\documentclass{article}
\usepackage{arxiv}
\usepackage[utf8]{inputenc} 
\usepackage[T1]{fontenc}    
\usepackage{hyperref}       
\usepackage{url}            
\usepackage{booktabs}       
\usepackage{amsfonts}       
\usepackage{nicefrac}       
\usepackage{microtype}      
\usepackage{lipsum}
\usepackage{graphicx}
\graphicspath{ {./images/} }
\usepackage{subcaption}  
\usepackage{float}        
\usepackage{amssymb}
\usepackage[export]{adjustbox}
\usepackage{xcolor}
\usepackage{tcolorbox}
\usepackage{amsmath}
\usepackage{enumitem}
\usepackage{framed}
\usepackage{booktabs}  
\usepackage{tocloft}
\usepackage{multirow}
\usepackage{threeparttable}

\title{Homophily-induced Emergence of Biased Structures in LLM-based Multi-Agent AI Systems
}

\author{
  Aliakbar Mehdizadeh  \\
  Department of Communication \\
  University of California, Davis \\
  \texttt{amehdizadeh@ucdavis.edu} \\
  \And
  Martin Hilbert \thanks{Corresponding author: \texttt{hilbert@ucdavis.edu}} \\
  Department of Communication \\
  University of California, Davis\\
  \texttt{hilbert@ucdavis.edu}
}

\begin{document}
\maketitle
\begin{abstract}
This study examines how interactions among artificially intelligent (AI) agents, guided by large language models (LLMs), drive the evolution of collective network structures. We ask LLM-driven agents to grow a network by informing them about current link constellations. Our observations confirm that agents consistently apply a preferential attachment mechanism, favoring connections to nodes with higher degrees. We systematically solicited more than a million decisions from four different LLMs, including Gemini, ChatGPT, Llama, and Claude. When social attributes such as age, gender, religion, and political orientation are incorporated, the resulting networks exhibit heightened assortativity, leading to the formation of distinct homophilic communities. This significantly alters the network topology from what would be expected under a pure preferential attachment model alone. Political and religious attributes most significantly fragment the collective, fostering polarized subgroups, while age and gender yield more gradual structural shifts. Strikingly, LLMs also reveal asymmetric patterns in heterophilous ties, suggesting embedded directional biases reflective of societal norms. As autonomous AI agents increasingly shape the architecture of online systems, these findings contribute to how algorithmic choices of generative AI collectives not only reshape network topology, but offer critical insights into how AI-driven systems co-evolve and self-organize.
\end{abstract}

\keywords{Large Language Models \and Multi-Agent Systems \and Network Growth \and Homophily}


\section{Introduction}

As artificial intelligence (AI) is increasingly given agentic autonomy, where algorithms reason, plan, converse, collaborate, cooperate, and compete, the very fabric of our society is being redefined by algorithmic choices. Agentic robotic trains, vacuum cleaner bots, and warehouse robots, have long been joined by less embodied AI systems that make autonomous decisions in recommending content on social media, scalping bots on the stock market, and editing bots that maintain websites. Most recently, this group has been expanded and empowered by a new generation of conversational generative AI, which is leading to a new wave of agentic AI. AI agents are becoming ever more present and influential social actors. This evolution has given rise to a new kind of social dynamic that goes beyond human-machine interactions by creating intricate patterns of machine–machine dynamics, which are necessary to understand  to maintain the ability to predict and steer the collective behavior of today's interconnected systems~\cite{tsvetkova2024new,rahwan2019machine}.

\subsection{The AI Agent Paradigm}

The AI agent paradigm is quite different from historic research on human–machine interactions, which has predominantly examined how humans perceive and engage with machines~\cite{nass1994computers, reeves1996media, lee2004trust}, which reflects an underlying assumption that machines function primarily as passive tools or assistants within human-centered decision-making processes. This orientation was rooted in the technological limitations of earlier systems, which lacked the autonomy or cognitive sophistication to act beyond pre-programmed instructions~\cite{parasuraman2000model}. Consequently, the burden of interpretation, judgment, and agency remained with human users, while passive machines served supportive or facilitative roles. However, with the advent of generative artificial intelligence systems, particularly large language models, this dynamic is undergoing a fundamental transformation~\cite{bommasani2021opportunities,hilbert2025ai}. These systems are not only increasingly capable of processing complex information, but, since they are generative, they can generate outcomes to decisions, which gives them an inherent level of autonomy, ~\cite{amodei2016concrete, bubeck2023sparks,brown2020language}, and, hence, agency. In short: one can ask ChatGPT "what would you do in that situation?" and it will generate an answer: generative AIs can, per definition, make decisions. 

"An autonomous agent is a system situated within and a part of an environment that senses that environment and acts on it, over time, in pursuit of its own agenda and so as to effect what it senses in the future"~\cite{franklin1996agent}. While most default generative AIs, including Large Language Models (LLMs) like ChatGPT, do not check all of these boxes, it is straightforward to convert a generative AI that can sense the environment and act on it (e.g. connect it to the web or tools through an API) and pursue an agenda (given to the LLM by a prompt, and, as such, turn it into an "autonomous agent"~\cite{maes1995artificial}, simply by soliciting it to generate decisions. The most common application is to link such decision capacity to the ability to access databases, in company intern environments, web searches, in research environments, or tool-use through APIs, in commercial environments ~\cite{wang2024survey}. Generative AI can also be asked to generate a workflow that outlines an entire path of action, which can be built to constantly and autonomously react to an environmental change, which can then be used as a blueprint for further action by the same AI, by recursively feeding the output into the generative AI as prompts~\cite{hilbert2025ai,wang2024survey}. 

While autonomous tool use  (in the form of APIs and databases) is the most common AI agent application, it is just as straightforward to connect an AI agent to other AI agents. In the words of NVIDIA's CEO Jensen Huang: "agents... can reason with each other and collaborate with each other and... go find other agents that can... work together and solve... problems"~\cite{cnet2024dreamforce} The resulting multi-agent systems based on generative AI have already found applications in software development~\cite{qian2023chatdev}, education~\cite{jiang2024ai,guo2024using}, mechanical engineering~\cite{ni2024mechagents},  power control~\cite{zhen2025online}, religious phenomenology ~\cite{ayrey2024llmtheism}, medical evaluation~\cite{fan2024ai}, programming~\cite{hong2023metagpt}, newspaper article writing~\cite{lin2025hybrid}, and other fields~\cite{wang2024survey}. Recent work has begun to explore whether these agentic systems can reproduce fundamental network properties, such as the emergence of scale-free networks through preferential attachment~\cite{de2023emergence}. Our study builds on this by investigating a different, but equally crucial, driver of human social networks: homophily, or the tendency for individuals to connect with similar others. 

As soon as more than two AI agents connect, networks emerge, which opens up a broad research agenda that combines social network analysis and multi-agent systems~\cite{hilbert2025module4}. Social networks are composed of interdependent and interacting agents (humans or machines), where the collective outcomes often cannot be predicted solely based on the traits and behavior of independent and distinct parts, given the nature and dynamic role of their links~\cite{bianconi2023complex}. Understanding the mechanisms that drive the formation and evolution of  abstract networks has been a cornerstone of research in network science~\cite{barabasi1999emergence,newman2003structure,jackson2008social}. The ability of Large Language Models (LLMs) to generate decision processes ~\cite{vaswani2017attention, bommasani2021opportunities,cui2024ai,burton2024large}. and their nuanced, but tractable reasoning based on contextual, incomplete or constrained information~\cite{hagendorff2023human,suri2024large}, allows for a rich research agenda that can provide valuable insights into the nature and consequences of the myriad of networks currently created by multi-agent AI systems.

\subsection{Exploring Agentic Network Generation}

The emerging field of AI-agent network science includes studies on how information diffuses on existing network structures~\cite{zhang2025llm}, but also on how autonomous agents create network structures. In this study, we focus on agentic network generation. We enter this far-reaching research agenda with the question of how unassuming preferences of autonomous AI agents on the micro-level can influence network structures on the macro-level. We systematically focus on one of the most common motivators of micro-behaviors in networks (namely, homophily) and one of the most common macro-network structures (namely, preferential attachment). These two well-known characteristics of social networks are not only well understood theoretically, but also very well echo network dynamics of human societies, and, hence, we can expect that AI that learned from human patterns, will exhibit a tendency to simulate these network characteristics. Starting from a small, fully connected initial graph, we ask AI agents to grow the network iteratively as new nodes query an LLM to determine their connections. We follow two rudimentary assumptions: 

\begin{itemize}
    \item one assumption is that the popularity of social actors in the existing network is of interest to the joining agent, which, in human networks, often result in a \textbf{preferential attachment} strategy, i.e. the 'rich-get-richer' in terms of connections~
    \cite{udny1925mathematical,simon1955class,price1976general,barabasi1999emergence};
    \item another assumption is that the characteristics of the node attributes of both the existing and joining nodes are of interest to the joining agent, which, in human networks, often results in \textbf{homophily}, i.e. 'birds-of-a-feather-flock-together'~\cite{lazarsfeld1954friendship,mcpherson2001birds}.
\end{itemize}

This simple setup allows us to create a transparent framework that lends itself to study the essence of the influence of informed, context-specific decision making of autonomous decision-making of AI agents on the emergent multi-agent network topology. By varying parameters such as the node attributes, the size of accessible subgroups and the number of connections formed per node, the framework systematically explores how constrained decision processes shape the structural and emergent properties of the network. This approach not only improves the understanding of societal network evolution that is increasingly influenced by agentic choices, but also contributes to the larger discourse on fairness and bias in AI systems~\cite{ferrara2023fairness,li2023survey}. This is because the result is basically an algorithm audit of how LLMs make network connections~\cite{bandy2021problematic,sandvig2014auditing}. 


\section{Methodology}

\begin{figure}[h!]
    \centering
    \includegraphics[width=\linewidth]{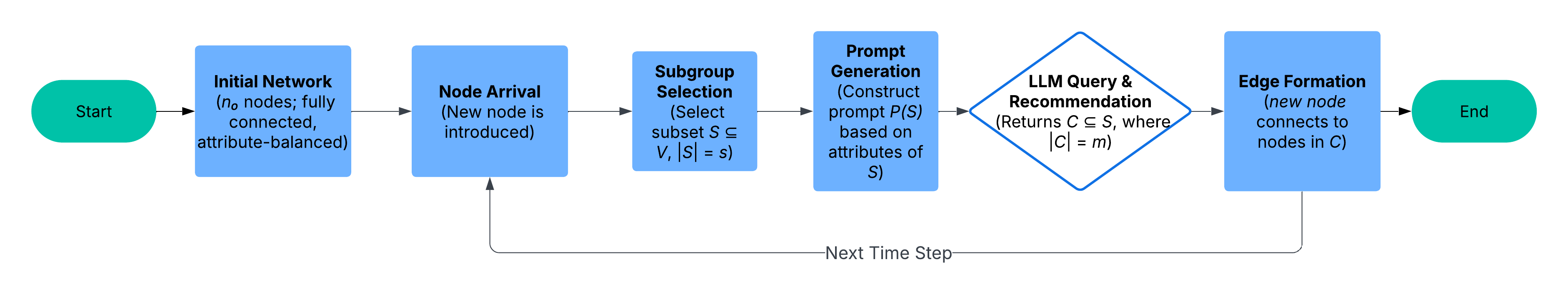}
    \caption{Flowchart of the node addition process in the LLM-driven evolving network model.}
    \label{fig:flowchart}
\end{figure}

Our chosen framework integrates decision-making processes informed by subgroup characteristics to dynamically simulate network growth. Key aspects of the methodology, including initialization, node addition process, prompt design, and experimental parameters, are detailed below, as depicted in Fig.~\ref{fig:flowchart} The procedure begins with an initial complete undirected graph, which serves as the seed network for the evolution process. New nodes are introduced sequentially, and their undirected connections are determined based on decisions made by an LLM. Mirroring common social processes, these decisions incorporate partial information on the existing network, such as degrees and attribute of the nodes. 

For any given simulation, we investigate one social attribute in isolation. Therefore, each node $v \in V$ is assigned a single categorical attribute, $A_v$, from one of the dimensions listed in Table~\ref{table:binary_attributes} (e.g., in one set of simulations, $A_v$ is ``Political Orientation'' with the 
value ``Liberal'' or ``Conservative''; in another set, $A_v$ is ``Gender'' with  the value ``Male'' or ``Female''). In all scenarios, the node's degree, $\deg(v)$, is also included as a structural attribute. The node addition procedure can be formalized as follows:
\begin{itemize}
    \item \textit{Subgroup Selection:} Select a random subset $S \subseteq V$ of size $s$ from the existing network. This subset $S$ represents the neighborhood of accessible nodes for the new node.
    \item \textit{Prompt Design:} Generate a prompt that summarizes the attribute of the nodes in $S$. The prompt is formulated as a vector of characteristics. The prompt asks the LLM to select exactly \( m \) members from this subset to form its initial connections. See Sec.~\ref{subsec:representative_prompt} for a representative prompt. 
    \item \textit{LLM Query:} Through a stateless API call, an LLM evaluates the prompt and outputs a set of recommended nodes, denoted as $C = \{c_1, c_2, \dots, c_m\}$, where $m$ is the number of connections the new node should form ( $C \subseteq S$).
    \item \textit{Connection Formation:} The new node $v_{new}$ establishes $m$ edges with the recommended nodes in $C$, adding $m$ edges to the network: $E' = E \cup \{ (v_{new}, c_i) \mid c_i \in C \}$.
\end{itemize}

\noindent The probability of selecting a node $v_i$ from $S$ for connection can be formulated as:

\[
\Pr(v_i \mid S) = f\left( a_{\text{new}}, \deg(v_{\textit{new}}), \{a_j\}_{v_j \in S}, \{\deg(v_j)\}_{v_j \in S} \right),
\]

where $\deg(v_i)$ denotes the degree of node $v_i$, $\{a_j\}_{v_j \in S}$, $\{\deg(v_j)\}_{v_j \in S}$ represent the attributes and degrees of all nodes in $S$, and $a_{\text{new}}$, and $ \deg(v_{\textit{new}})$  denotes the attribute and degree vector of the newly introduced node, respectively. The function $f(\cdot)$ is a stochastic decision mechanism embedded within the architecture of the LLM under study, capturing the relative importance of the attribute of the nodes, the degree distribution and the overall structure of the network to determine the probabilities of attachment.

\subsection{Model Specification}
\label{subsec:model_spec}

Let the initial network be denoted by $G_0 = (V_0, E_0)$, where $|V_0|$ specifies the number of nodes in the fully connected attributed graph that serves as the starting point of the simulation. To ensure impartiality with respect to the attribute under investigation, the starting network is designed to be complete, attributed, and balanced, mitigating any predisposition toward a particular attribute category. 

\subsubsection{Network Generation}
\label{subsec:network_generation}

The network evolves via the sequential introduction of new nodes until the network reaches a total size of $|V_f|$ nodes. The subgroup size, represented by ( s ), specifies the number of existing nodes evaluated when a new node is introduced, thereby shaping the information available for attachment decisions. This models a realistic scenario of partial information, where agents are not aware of all other participants in the network. This contrasts with other models, such as that of De Marzo et al~\cite{de2023emergence}, where agents are presented with a list of all existing users, more closely mirroring the classic Barabási-Albert model's assumption of global information. The connections per node, denoted as ( m ), specifies the number of edges each new node establishes within the selected subgroup. A pre-trained Large Language Model (LLM) is queried at each time step to evaluate the attributes and structure of the nodes in $S$ and recommend $m$ nodes for connection, operating under specific configurations including parameters such as temperature, which influences the randomness and creativity of the model's outputs during decision-making processes. The parameter space explored in our experiments is as follows:

\begin{itemize}
    \item \textit{Initial network size}: $|V_0| = 4$ (a fully connected and attribute-balanced graph).
    \item \textit{Subgroup size}: $s \in \{10, 50\}$, representing small and large levels of network accessibility.
    \item \textit{Connections per node}: $m \in \{2, 3, 5\}$, varying the level of connectivity per new node.
    \item \textit{Target network size}: $|V_f| = 1500$ nodes.
    \item \textit{LLM models}: Gemini 1.5 Flash, GPT-4o Mini, Llama.4.Scout, Claude 3 Haiku.
\end{itemize}

\subsubsection{Representative Prompt}
\label{subsec:representative_prompt}

Prompts play a critical role in shaping the behavior of large language models (LLMs), functioning as interfaces through which task structure and contextual information are communicated. In this study, prompts are constructed to simulate social onboarding decisions, prompting the model to choose connections based on both demographic similarity (homophily) and structural cues (e.g., degree centrality). This approach draws from recent advances in prompt engineering, which demonstrate that even simple natural language formulations can reliably induce complex, context-sensitive behavior in LLMs~\cite{brown2020language, wei2022chain, zhou2022least}. By framing the task in familiar social terms, the prompt facilitates alignment with the model’s pretrained knowledge of human behavior, enabling the exploration of how such models internalize and reproduce social biases and heuristics in network formation~\cite{argyle2022out, schick2021self}.

The following is a representative prompt used to elicit attachment decisions from the language model during the simulation. This prompt is designed to reflect a typical onboarding scenario within a growing social network, where the model is asked to make link formation choices based on limited demographic and structural information. For example, the [Attribute] could be gender and Option 1 = male and Option 2 = female. In the degree-exclusive scenario, only the number of connections (i.e., node degrees) is provided to the LLMs, without any attribute or identity information.

\begin{framed}
\textit{You are a new member joining a growing social network. Your demographic attributes are as follows:}  
\textit{[Attribute]: Option 1}  

\textit{Below is a randomly selected subset of existing members, along with their number of connections and gender:}  

\begin{itemize}[noitemsep, topsep=0pt]
    \item \textbf{Member} \textit{"fefg"}: 7 connections; [Attribute]: Option 1
    \item \textbf{Member} \textit{"fthk"}: 2 connections; [Attribute]: Option 2
    \item \textbf{Member} \textit{"terq"}: 6 connections; [Attribute]: Option 2
    \item \dots
    \item \textbf{Member} \textit{"vklp"}: 4 connections; [Attribute]: Option 1
\end{itemize}

\textit{Select exactly \( m \) members from this subset to form your initial connections. Provide the selected member IDs as a comma-separated list, without any additional explanation.}
\end{framed}

The prompt shown above operationalizes the theoretical framework by mapping each variable in the formalization to a concrete, human-readable format interpretable by the LLM. The new node's attributes, \( a_{\text{new}} \), such as gender, are specified at the beginning of the prompt (``Your demographic attributes are\ldots''). The subset \( S \subseteq V \), representing the neighborhood of accessible nodes, is introduced as the ``randomly selected subset of existing members,'' where each member is associated with an attribute \( \{a_j\}_{v_j \in S} \) and degrees \( \{\deg(v_j)\}_{v_j \in S} \) (e.g., number of connections and gender). These correspond to the lines listing each member’s ID, degree, and demographic trait. The LLM’s task—selecting \( m \) members from this list directly corresponds to computing the set \( C \subseteq S \) based on the stochastic decision function \( f(\cdot) \). In this way, the prompt acts as a computational instantiation of the theoretical decision process, embedding the structure of \( G \), the characteristics of \( S \), and the attributes of the new node into a textual interface for probabilistic attachment.


\subsection{Preferential attachment}
\label{subsec:pref_attachment}

The basic dynamic of sequentially adding one node at a time follows common practice of the longstanding default generative mechanism for preferential attachment networks~\cite{yule1925ii,simon1955class,price1976general,barabasi1999emergence}. We analyze the degree distributions of four distinct node attachment strategies: Highest Degree Attachment, Preferential Attachment~\cite{barabasi1999emergence}, Partial Preferential Attachment~\cite{carletti2015preferential}, and LLM-Assisted Attachment. While each model represents a different strategy for network growth and evolution, comparing them allows us to do a comparative analysis that reveals deviations from well-understood attachment strategies:

\begin{itemize}
    \item \textit{Highest Degree Attachment}: For each new node, we randomly sample $s$ existing nodes and connect the new node to the $m$ nodes in this subsample with the highest degrees.    
    \item \textit{Preferential Attachment}: The classical model in which new nodes attach with a probability directly proportional to the degree of existing nodes.
    \item \textit{Partial Preferential Attachment}: New nodes receive a randomly sampled subset of existing nodes and attach to one within this subset with a probability proportional to its degree.
    \item \textit{LLM-assisted Attachment}: Large language models (LLMs) determine attachment decisions based on nodes attribute and degrees.

\end{itemize}

To assess the impact of different attachment strategies, we conducted a series of controlled experiments incorporating various node attributes. The baseline condition considered only degree-based attachment, while subsequent conditions included additional attributes in conjunction with degree. This methodological design allows us to systematically evaluate how the inclusion of additional node features influences network topology relative to the degree-only baseline. Each experiment was repeated 30 times to ensure statistical robustness, and the results were averaged to identify trends in network evolution. 


\subsection{Homophily}
\label{subsec:node_attributes}

To examine how LLM-based AI agents conceptualize the formation of social networks, we identify and analyze key attributes that influence attachment strategies within these systems. The formation of social ties is frequently influenced by the well-established principle of homophily, whereby individuals exhibit a preference for connecting with others who share similar traits~\cite{mcpherson2001birds, kossinets2009origins,khanam2023homophily}. To systematically model these patterns, we operationalize individual attributes by classifying them into binary categories~\cite{newman2003mixing}. This categorization facilitates the translation of qualitative social characteristics into structured variables that inform and constrain the network's growth dynamics. 
In order to identify relevant node attributes among a myriad of possible ones, we walk the talk of the agentic paradigm and opt for a mixed-method approach between more automated computational exploration and a more traditional literature review synthesis. To obtain expansive breadth, we consult 10 different foundational models about "the most prominent node attributes usually considered by humans when looking for similar others" (see Supporting Information~\ref{sec:S.I.1. Identifying Node Attributes}) and rank the ones most commonly identified by the AIs. We then focus with a traditional literature review to reduce this selection by half (see below, in this section), which leaves us with eight different node attributes. The result includes representative attributes from both groups of the longstanding distinction from homophily's founding paper, i.e. status homophily and value homophily, ~\cite{lazarsfeld1954friendship} as outlined in Table~\ref{table:binary_attributes}. Following status homophily, people associate based on more static shared social characteristics (both innate and acquired), and following value homophily, people associate based on less static shared beliefs and attitudes regardless of social status differences.

\begin{table}[h!]
\centering
\small
\renewcommand{\arraystretch}{1.2}
\begin{tabular}{@{} l l l l @{}}
\toprule
\textbf{Status vs. Value} & \textbf{Attribute} & \textbf{Category 1} & \textbf{Category 2} \\
\midrule
Status & Gender & Female & Male \\
Status & Education & College Degree & No Degree \\
Status & Ethnicity & Western & Non-Western \\
Status & Socio-economic Status & Professional & Working Class \\
Status & Age & $<$30 years & $\geq$30 years \\
Value & Religious Practice & Practicing & Non-practicing \\
Value & Political Orientation & Liberal & Conservative \\
Value & Interests\footnotemark & Active & Passive \\
\bottomrule
\end{tabular}
\vspace{10pt}
\caption{Binary attribute classifications used in network labeling.}
\label{table:binary_attributes}
\end{table}

\footnotetext{This attribute was excluded from the final analysis. See Section 2.3.8 for justification.}

\subsubsection{GENDER}

The literature agrees on the influence of gender on social tie formation and network topology ~\cite{brands2022theorizing}. A substantial body of research has examined the mechanisms through which gender disparities are reproduced via the structural and dynamic processes that underpin social networks~\cite{belliveau2005blind, ding2013gender, smith2000mobilizing}. Foundational theoretical frameworks such as homophily have played a critical role in explaining how gender similarity fosters the formation and reinforcement of interpersonal ties~\cite{mcpherson2001birds, ibarra1992homophily}. In contrast, the concept of heterophily explores the conditions under which cross-gender ties are formed and sustained, often in contexts marked by power asymmetries, role expectations, or status differentials~\cite{blau1977inequality, granovetter1973strength, faris2011status, bailey2025cross}. Parallel lines of inquiry have focused on gendered access to social capital, investigating how differences in network composition and tie quality may produce unequal access to information, influence, or institutional resources~\cite{burt2003social, lin2002social, podolny1997resources, moore1990structural}. These structural asymmetries are further evidenced in analyses of network-level indicators, including size, density, centrality, and composition, which frequently reveal systematic gender differences in how individuals are embedded within broader social systems~\cite{brass1985men, ibarra1997paving}. With the emergence of machine learning, and large language models (LLMs), a new research frontier has opened for examining how such models internalize, reflect, or potentially amplify gendered social patterns. Recent studies suggest that models that are trained on large-scale web data often encode and reproduce gender stereotypes, which may shape how they represent social roles, occupations, and interpersonal dynamics~\cite{bolukbasi2016man, zhao2018gender, sheng2019woman, kaneko2024evaluating,wan2023kelly}. As these systems become integrated into decision-making pipelines that influence hiring, recommendations, and content generation, understanding their gendered representations of social structure becomes increasingly urgent~\cite{nadeem2020stereoset, kotek2023gender}.

\subsubsection{EDUCATION}

Education level plays a significant role in shaping the formation and structure of social ties, often influencing patterns of homophily~\cite{mcpherson2001birds, dimaggio1985cultural, mare1991five}. Individuals with similar educational backgrounds are more likely to form connections, leading to the development of stratified social networks that reinforce existing social hierarchies. However, other studies suggest that individuals with higher educational attainment tend to cultivate larger and more diverse social networks~\cite{dika2002applications}. This is particularly true in educational settings, where higher education provides opportunities for building social capital through the development of social skills, as well as the formation of friendships and professional connections~\cite{helliwell1999education}.

\subsubsection{ETHNICITY}

Demographic and cultural attributes such as religion, social class, ethnicity, and personal interests have long been recognized as fundamental axes of social tie formation and network stratification~\cite{mcpherson2001birds,blau1977inequality,fischer1982dwell,lamont1992money}. Classic research has documented the strong effects of ethnic homophily in shaping both offline and online communities, influencing the density, cohesion, and informational flow within groups~\cite{huckfeldt1995citizens,de2011tie}. Similarly, class and educational background often correlate with network composition and access to social capital, with stratified ties influencing opportunities and mobility across the life course~\cite{lin2002social, zweigenhaft2006diversity, granovetter1973strength}. Shared interests—whether cultural, political, or recreational—further shape community formation and tie strength, providing both bonding and bridging capital~\cite{putnam2000bowling}. Studies show that machines can reproduce societal biases and stereotypes tied to religion, race, class, and educational background~\cite{abid2021persistent,mehrabi2021survey,garg2018word}. For example, evaluations reveal that models often associate specific religious or ethnic identities with violent or negative terms~\cite{abid2021persistent}.

\subsubsection{SOCIOECONOMIC}

Socioeconomic status significantly influences the formation and maintenance of social ties, often acting as a barrier or facilitator to network diversity and reach~\cite{lin2000inequality,gould2016growing, small2009unanticipated, han2015social}. Individuals with similar socioeconomic status are more likely to interact and form relationships, contributing to homophilous networks that limit cross-class connections~\cite{mcpherson2001birds}. Moreover, lower socioeconomic status such as working-class individuals may face structural constraints such as limited access to resources or geographic immobility that hinder the development of broader social networks~\cite{small2009unanticipated}. Conversely, higher socioeconomic status such as professionals is often associated with access to more heterogeneous and resource-rich networks, enabling the accumulation of social capital~\cite{lin2000inequality}.

\subsubsection{AGE}

Similar to gender, the literature identifies age as a fundamental demographic characteristic that significantly influences the formation of social ties across various social contexts. Individuals often demonstrate a tendency to form stronger and more frequent ties with those within similar age ranges, which is due to both status considerations, such as generational cohorts, and value considerations, such as shared life experiences and common interests ~\cite{mcpherson2001birds,brashears2008gender,ertug2022does}. Conversely, the formation and maintenance of heterophilous ties across age groups are also crucial for intergenerational understanding, the diffusion of innovation, and the transfer of knowledge and experience ~\cite{granovetter1973strength,franken2025unstable}. Research on social capital explores whether age-related patterns of tie formation result in differential access to resources and support, with younger and older individuals potentially leveraging different types of networks for varying needs ~\cite{uhlenberg2004age,fuhse2024networks,weiss2022life}. Furthermore, analyses of network structure investigate how age influences the size, density, and composition of individuals' social networks, with life cycle stages potentially shaping the diversity and nature of their connections ~\cite{de2011tie,jeroense2024similarity}. 


\subsubsection{RELIGIOUS PARTICIPATION}

While Lazarsfeld and Merton \cite{lazarsfeld1954friendship} saw religion as socially-ascribed 'status', we implemented it as a less static, and more attitude-, and belief-based value, asking it about the active commitment of practicing religion or not. Religious participation has long been recognized as a significant factor in the formation and maintenance of social ties, often fostering dense, supportive, and enduring social networks~\cite{lim2010religion, putnam2000bowling, cheadle2012friendship}. Religious settings offer regular opportunities for face-to-face interaction, shared rituals, and collective identity, all of which strengthen bonding social capital. Moreover, religious communities frequently facilitate bridging ties across socioeconomic or ethnic boundaries, though the extent of this inclusiveness can vary widely depending on denominational and geographic context~\cite{wuthnow2002religious}. These networks not only offer emotional and spiritual support but also serve as important conduits for civic engagement and collective action~\cite{merino2014social,lewis2013religion}.

\subsubsection{POLITICAL ORIENTATION}

Political homophily, the tendency to connect with those who share similar political views, is also a well-documented phenomenon in the literature ~\cite{mcpherson2001birds, barbera2015tweeting,halberstam2016homophily,de2011tie}. It has become a hot topic of research, since in politically polarized societies, this can lead to the formation of echo chambers and filter bubbles, where individuals are primarily exposed to like-minded perspectives, reinforcing existing beliefs and potentially exacerbating societal divisions ~\cite{sunstein2009going, pariser2011filter}. Moreover, the broader political environment, including the nature of the regime (democratic vs. authoritarian), the level of political freedom, and the presence of social movements, can significantly influence who individuals connect with and the types of networks they build. In authoritarian contexts, for instance, individuals may be more cautious in forming ties, particularly those that could be perceived as politically oppositional, leading to the development of  tightly knit networks for dissent and mobilization ~\cite{granovetter1973strength,chong2014collective}. Recent research has also begun to explore how large language models (LLMs) internalize and reproduce political ideologies and biases present in their training data~\cite{hartvigsen2022toxigen, feng2023pretraining, argyle2023out, gallegos2024bias}. Studies indicate that LLMs often reflect the ideological slant of their data sources, which can influence how they respond to politically charged prompts or frame contentious issues. Some models exhibit measurable political leanings, raising concerns about how their outputs may inadvertently reinforce polarization or shape public discourse in biased ways~\cite{bang2024measuring,yang2024unpacking,fulay2024relationship}.

\subsubsection{INTEREST}

Shared interests are a powerful driver of social tie formation, particularly in contexts where individuals are free to associate based on preference rather than obligation~\cite{ kossinets2006empirical, uzzi2005build, rivera2010dynamics}. Common hobbies, values, or intellectual pursuits foster repeated interaction, mutual understanding, and trust, which in turn reinforce social cohesion. Online platforms, in particular, have amplified the role of interest-based affiliations by allowing individuals to form communities beyond geographic and demographic constraints~\cite{ellison2007benefits}. These interest-based ties often bridge diverse backgrounds, enabling the formation of weak ties that serve as conduits for novel information and opportunities~\cite{granovetter1973strength}. 

Initially, we included 'Interest' as an attribute, operationalized with a binary distinction between 'Active Engagement' and 'Passive Engagement'. However, this classification proved problematic during our analysis, yielding highly inconsistent mixing patterns across the different LLMs. Unlike the other attributes, 'Interest' is inherently high-dimensional and fragmented, and we found that this single binary variable was insufficient to capture its complexity. Given these issues with construct validity and the resulting ambiguity, we excluded this attribute from our final aggregate analysis to ensure the robustness of our comparisons. A more detailed analysis is available in Table~\ref{sec:assortativity_table} in the appendix.


\subsection{Measures}
\label{subsec:measures}
We compare key network properties across different experimental conditions, including degree distribution, clustering coefficient, average path length, degree assortativity, and attribute assortativity. The latter two are our main measures~\cite{newman2003mixing,newman2002assortative}. Attribute assortativity measures the tendency of nodes to connect with others sharing similar attributes (like age or gender), while degree assortativity measures the tendency of nodes to connect with others having similar numbers of connections; both are quantified using the assortativity coefficient, $r_{att}$ and $r_{deg}$, which ranges from -1 (perfect disassortativity) to +1 (perfect assortativity), with attribute assortativity directly measuring homophily based on node traits and degree assortativity revealing network structural patterns that can reinforce or counteract homophilic tendencies.

The general formula for attribute assortativity, which measures the correlation between scalar properties (like age) of connected vertices, is given by the Pearson coefficient~\cite{newman2003mixing}:

\[
r = \frac{1}{\sigma_a \sigma_b} \sum_{xy} xy \left( e_{xy} - a_x b_y \right)
\]

Here, $e_{xy}$ is the fraction of all edges that connect vertices with attribute values $x$ and $y$, while $a_x$ and $b_y$ are the fractions of edges that start or end at vertices with those respective values. A special, but important, case is degree assortativity, which measures the tendency of vertices to connect to others with a similar number of connections. The formula is a specific application of the above principle~\cite{newman2003mixing}:

\[
r = \frac{1}{\sigma_q^2} \sum_{jk} jk \left( e_{jk} - q_j q_k \right)
\]

In this equation, $j$ and $k$ represent the degrees of the connected vertices, $e_{jk}$ is the joint probability that an edge connects vertices of degree $j$ and $k$, and $q_k$ represents the excess degree distribution, the probability that a vertex at the end of a randomly chosen edge has degree $k$. For both measures, a value of $r > 0$ indicates assortativity, while $r < 0$ indicates disassortativity. For a full discussion on derivation and analysis, see Newman~\cite{newman2003mixing}. Note that we do not directly calculate these metrics ourselves; instead, we use the \texttt{networkx} library for their computation.

We also systematically assess formation rates in networks generated by large language models (LLMs). This focuses on the relative frequencies of connections between nodes categorized by our binary attributes. It captures the extent to which LLM-generated links reflect attribute assortativity, the tendency for nodes with similar categorical attributes to connect more frequently than by chance. This notion parallels the concept of homophily in social networks and is formalized in network science as assortative mixing by discrete node attributes. Through this, we can also examine whether edge formation between categories is balanced, for example, whether female nodes tend to initiate more connections with male nodes or vice versa. Such asymmetries in cross-group link formation can reveal directional biases in LLM-generated relational patterns and further inform our understanding of attribute-based mixing~\cite{newman2003mixing, mcpherson2001birds}.


\section{Results and Discussion}

\begin{figure}[h!]
    \centering
    \begin{subfigure}[b]{0.45\textwidth}
        \centering
        \includegraphics[width=\textwidth]{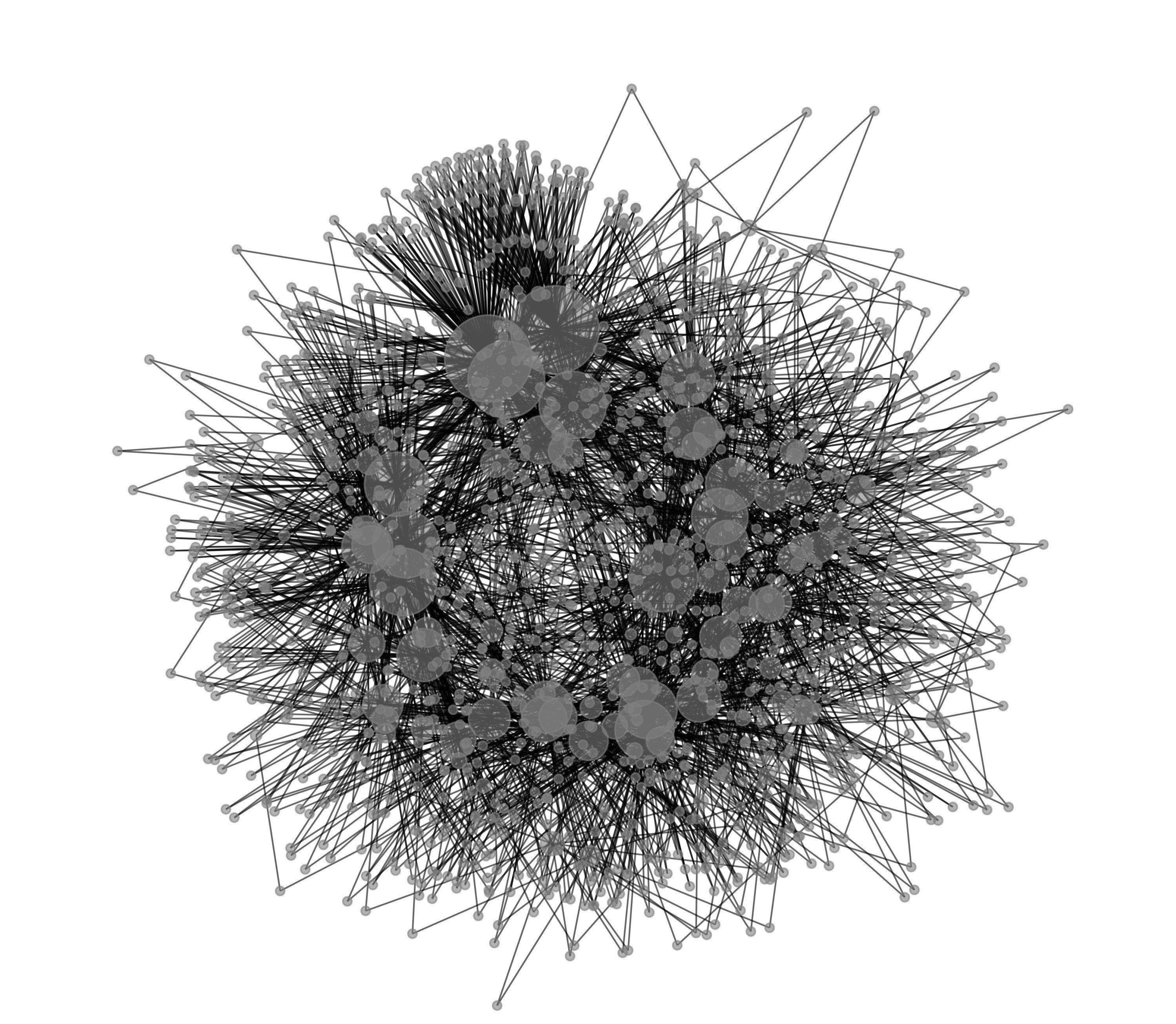}
        \caption{Degree-Exclusive (No attribute)}
        \label{fig:network_a}
    \end{subfigure}
    \hfill
    \begin{subfigure}[b]{0.45\textwidth}
        \centering
        \includegraphics[width=\textwidth]{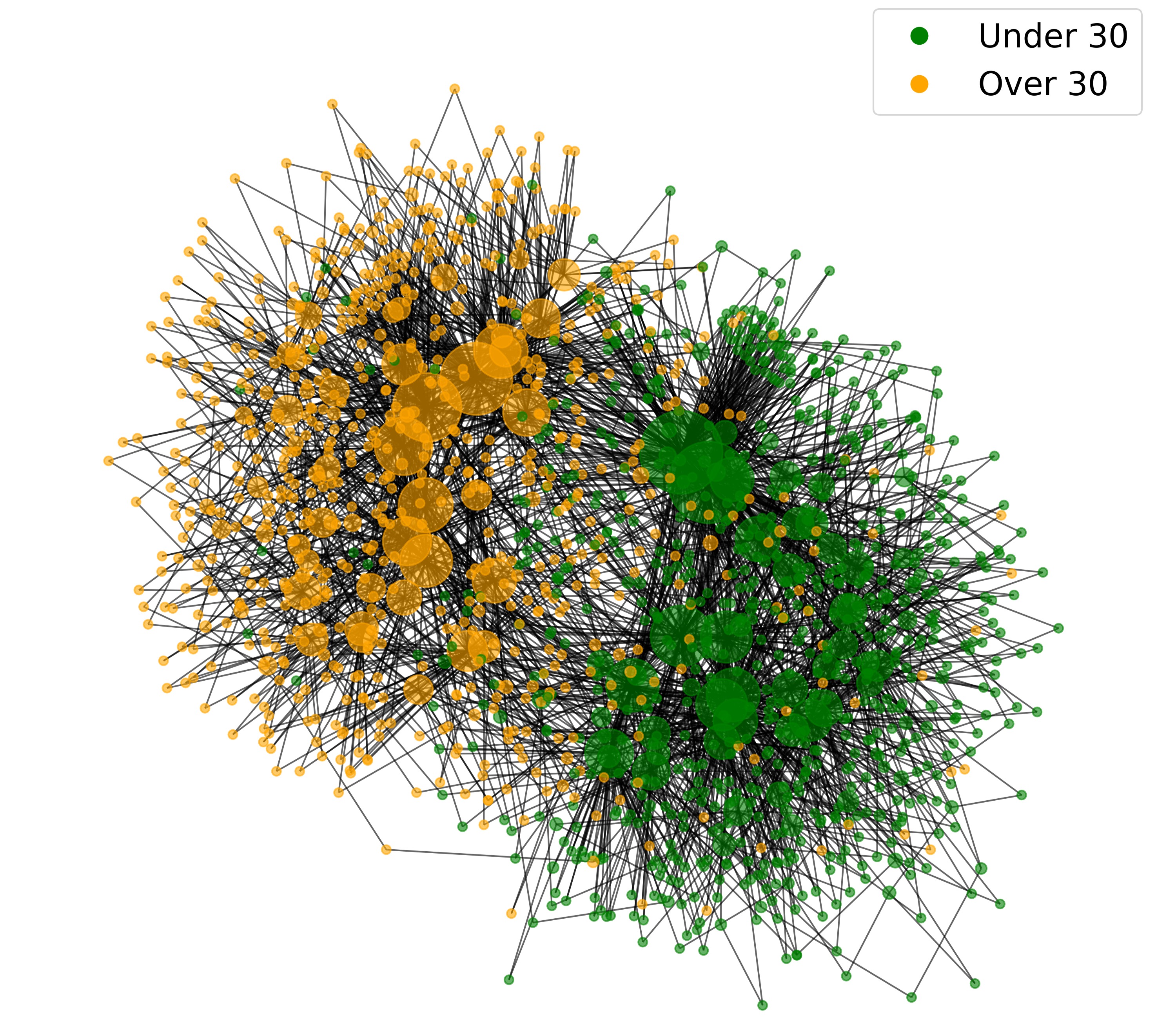}
        \caption{Age (Under 30 \& Over 30)}
        \label{fig:network_b}
    \end{subfigure}

    \vspace{0.5cm}

    \begin{subfigure}[b]{0.45\textwidth}
        \centering
        \includegraphics[width=\textwidth]{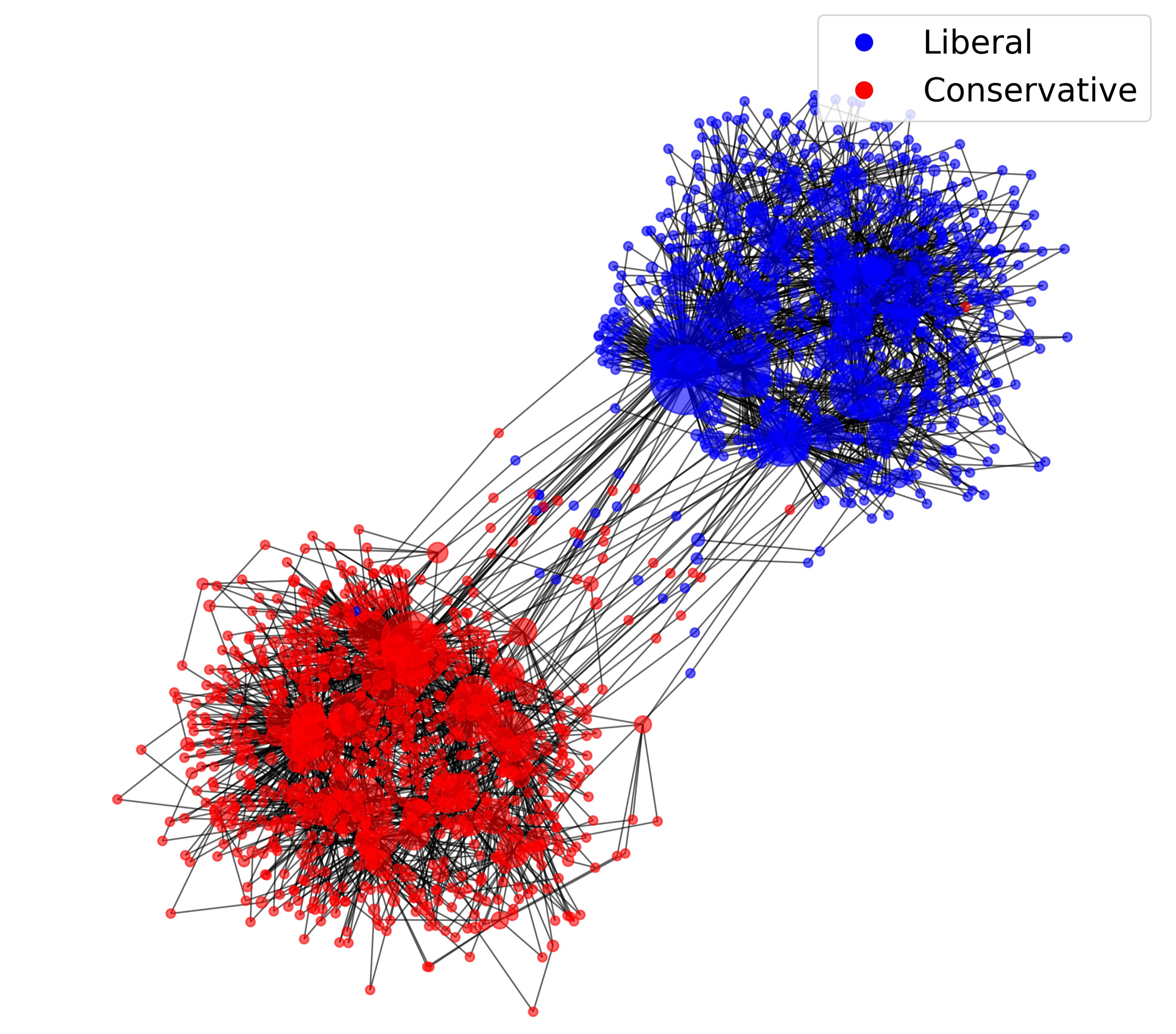}
        \caption{Political Orientation (Liberal \& Conservative)}
        \label{fig:network_c}
    \end{subfigure}
    \hfill
    \begin{subfigure}[b]{0.45\textwidth}
        \centering
        \includegraphics[width=\textwidth]{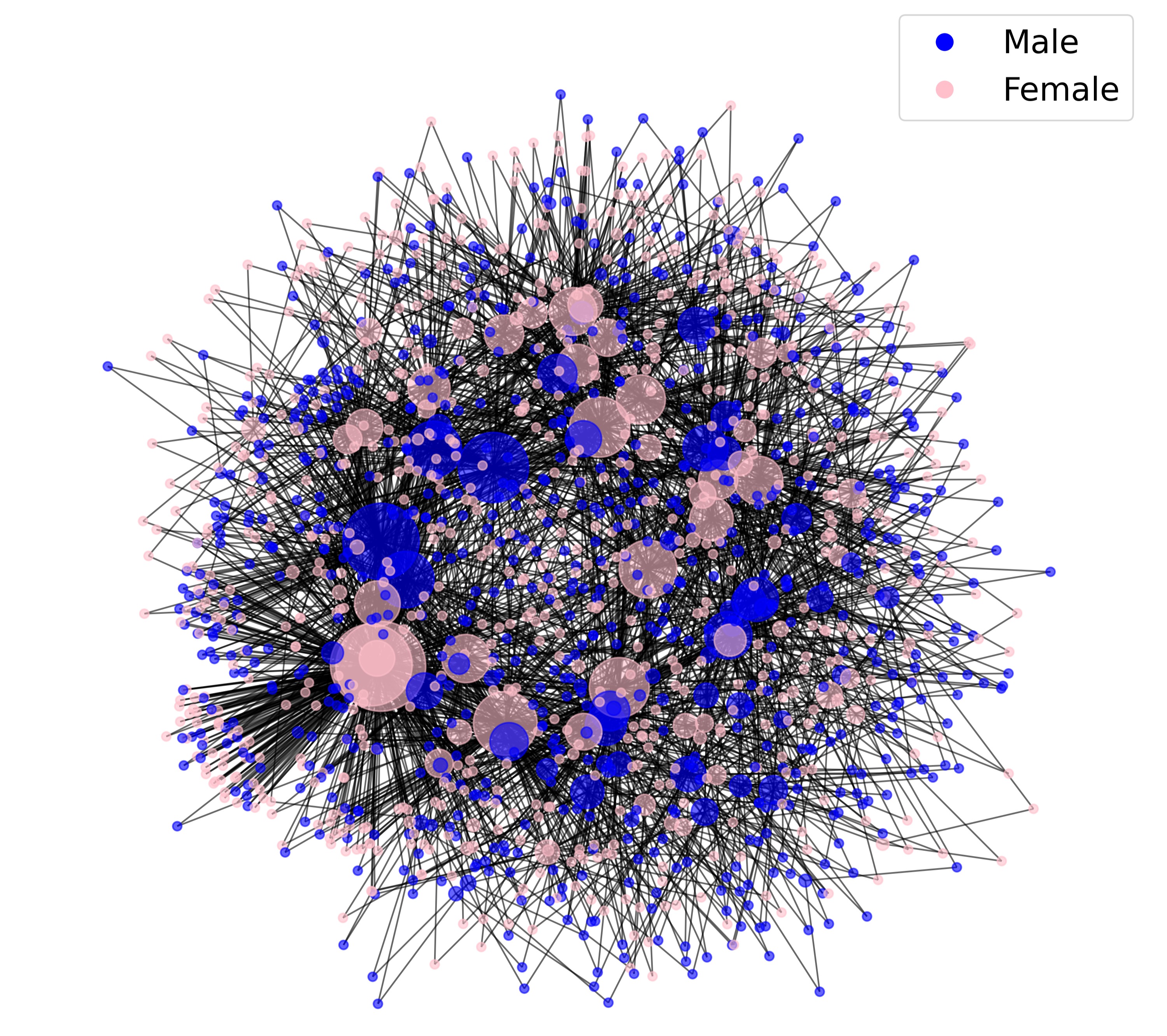}
        \caption{Gender (Male \& Female)}
        \label{fig:network_dd}
    \end{subfigure}

    \caption{Visualization of 4 networks, each illustrating different node attribute configurations. Node colors indicate attribute categories, while node sizes are proportional to degree. Each network comprises \( n = 1500 \) nodes, where each new node connects to \( m = 2 \) existing nodes selected from a random subsample of \( s = 50 \) nodes, based on recommendations generated by Gemini 1.5 Flash.}
    \label{fig:network_visualization}
\end{figure}

Figure~\ref{fig:network_visualization} visually illustrates our procedure for creating networks with LLMs, using a small selection of cases for demonstration purposes. While we see our baseline case of the degree-exclusive network in Figure\hyperref[fig:network_visualization]{~\ref*{fig:network_visualization} 
(a)}, which means nodes attach solely based on degree without attribute consideration, the other panels highlight how different attributes influence network topology. In Figure \hyperref[fig:network_visualization]{~\ref*{fig:network_visualization}(b)}, including age leads to clear clustering by age group, with younger and older nodes forming visibly distinct communities. Figure\hyperref[fig:network_visualization]{~\ref*{fig:network_visualization}(c)} shows strong assortativity in political orientation, with liberal and conservative nodes forming highly polarized, densely connected subgroups. Finally, Figure \hyperref[fig:network_visualization]{~\ref*{fig:network_visualization}(d)} reveals a more blended structure when gender is included. These contrasts illustrate how different types of attribute influence can reshape the emergent structure of the network.

\subsection{Basecase Degree Distribution Analysis of Attachment Strategies}

\begin{figure}[h!]
    \centering
    \includegraphics[width=0.9\textwidth]{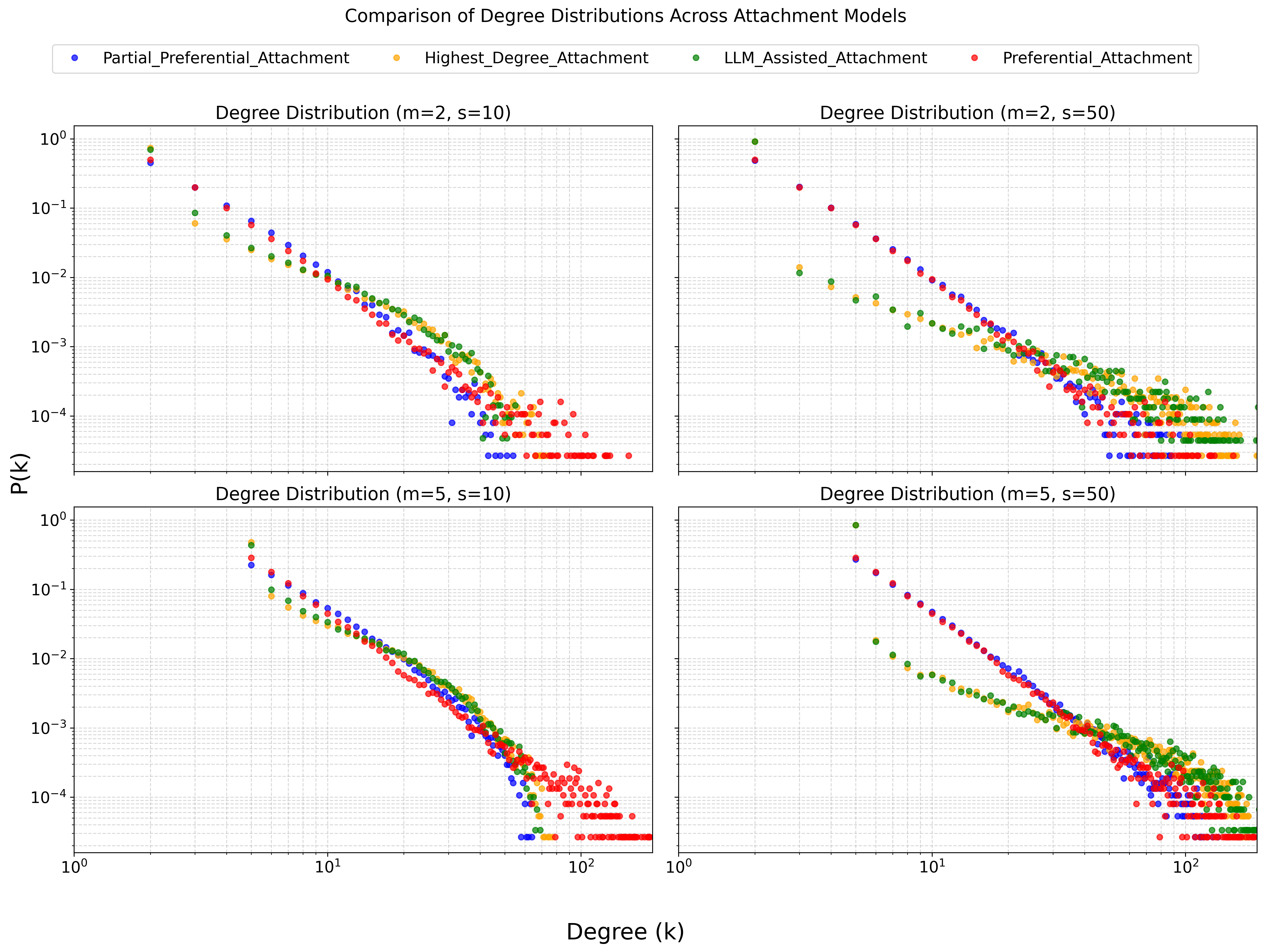}
    \caption{Comparison of degree distributions across different attachment strategies in network growth for $n = 1500$ nodes and two levels of connectivity ($m = 2$ and $m = 5$) and subsample sizes of $s = 10$ and $s = 50$. The strategies include: (i) highest-degree attachment , (ii) preferential attachment, (iii) partial preferential attachment, and (iv) LLM-assisted attachment generated using the Gemini 1.5 Flash model. The figure illustrates how each strategy shapes the emergent degree distribution. Results are averaged over 30 realizations}
    \label{fig:degree_distributions}
\end{figure}

Before diving deeper into our results, we test for the sensitivity of our base case to different network configurations, namely, different settings of our parameters $m$ (number of connections per node) and $s$ (number of nodes revealed to the joining node). Figure~\ref{fig:degree_distributions} presents the degree distributions of four distinct attachment strategies under varying parameter settings, specifically the number of edges added per new node $m$ and the sample size representing network accessibility $s$. We test four different attachment strategies: (i) \textit{highest-degree attachment}, which \emph{depends on $s$} since the highest-degree nodes are chosen from a random subsample of size $s$; (ii) \textit{preferential attachment}, which \emph{does not depend on $s$} because node selection probabilities are based on degrees across the entire network; (iii) \textit{partial preferential attachment}, where preferential attachment is applied within a subsample of size $s$; and (iv) \textit{LLM-assisted attachment} generated through language model predictions. These strategies are discussed in more detail in Section~\ref{subsec:pref_attachment}.

The parameters for subgroup size ($s$) and connections per node ($m$) were chosen to systematically explore a range of network growth conditions and identify a regime that clearly reveals the distinct behaviors of the LLM agents. The initial values for subgroup size, $s \in \{10, 50\}$, were selected to represent small and large levels of network accessibility, respectively. This allows for a comparison between a highly constrained information environment ($s=10$) and a more information-rich one ($s=50$). Similarly, the values for connections per node, $m \in \{2, 5\}$, were chosen to model varying levels of initial connectivity for each new agent.

The figure comprises four scatter plots, each depicting the degree distribution $P(k)$ as a function of degree $k$ on a logarithmic scale, for parameter combinations $m=[2,5]$ and $s=[10,50]$. Across all configurations, the degree distributions exhibit a characteristic power-law-like behavior, with $P(k)$ decreasing as $k$ increases, suggestive of scale-free network properties. However, definitive confirmation of a true power-law distribution, and thus a scale-free topology, typically requires the analysis of significantly larger networks (e.g., millions of nodes) to reliably discriminate from other heavy-tailed distributions like exponential networks. Nonetheless, notable differences emerge among the strategies, influenced by the parameters $m$ and $s$.

For smaller sample sizes ($s=10$), as observed in the left-hand plots, the degree distributions of all models display a steeper decline in the tail (higher ($k$)). This suggests a reduced likelihood of high-degree nodes due to limited network information access, which constrains the attachment options. The Partial Preferential Attachment strategies and the Preferential Attachment show similar distributions, with a dense clustering of points at lower degrees and a sharper drop-off at the tail. When the sample size of revealed nodes increases ($s=50$), as shown in the right-hand plots, the differences between models become more pronounced. The Preferential Attachment and Partial Preferential Attachment continue to show a steep decline in $P(k)$, but the tail extends further, reflecting a greater chance of high-degree nodes due to increased access to network information. 

The Highest Degree Attachment and LLM-Assisted Attachment, however, demonstrate a more gradual decline in the tail, particularly for $m=5$ and $s=50$. This suggests that these mechanisms are more effective at producing high-degree nodes under conditions of greater network visibility. This observation notes that the LLM-assisted closely resembles the Highest Degree Attachment in its degree distribution, rather than the classical Preferential Attachment: LLMs seem to either follow a reward function that motivates them to connect to the most popular others or simply orient themselves on the only thing they were given, which, in this case, is the number of links. The parameter $m$ also influences the distributions. For $m=5$ (bottom plots), the overall density of points increases compared to $m=2$ (top plots), reflecting the addition of more edges per node, which enhances connectivity and shifts the distribution toward higher degrees across all models. However, the effect of $s$ remains dominant in differentiating the models, as larger sample sizes amplify the distinct attachment behaviors.

To further validate these findings, the experiment was repeated using additional LLMs, including ChatGPT-4o Mini, Claude 3.0 Haiku, and Llama-4-Scout, all employing similar attachment strategies. We observe that with Gemini, the LLM-assisted model follows a highest-degree attachment pattern most strongly, as reflected in the Figure.~\ref{fig:degree_distributions}. The degree-exclusive models exhibit very close values of \(r_{\text{deg}}\), indicating that incoming nodes prioritize their connection strategy similarly across these models, see Figure.~\ref{fig:degree_assortativity}. A close similarity in \(r_{\text{deg}}\) values suggests that the attachment strategies employed by these LLMs are nearly identical. Furthermore, in the Figure~\ref{fig:model_follow_HDA}, we compare the degree distributions of the four models for networks with 2000 nodes, providing an additional dimension of similarity that corroborates their comparable attachment strategy. These results reinforce the notion that when provided solely with the number of links, regardless of the specific LLM or network size, the models tend to emulate highest-degree attachment strategies. 

Based on our observations, we fix \( m = 3 \) and \( s = 50 \) for the remaining simulations to operate in a regime where the network topology generated by LLM-assisted attachment meaningfully deviates from classical preferential attachment. The sample size parameter \( s \) plays a critical role in revealing distinctions between attachment strategies. As shown in the degree distributions, \( s = 50 \) amplifies the characteristic behaviors of each model, particularly highlighting how LLM-assisted attachment more closely resembles highest-degree attachment under conditions of increased visibility. We select \( m = 3 \) as a balanced intermediate setting: it introduces sufficient structural complexity while avoiding the excessive connectivity associated with larger \( m \), which can obscure differences in attachment behavior. Additionally, a smaller \( m \) places greater pressure on the model to make deliberate attachment decisions, effectively forcing a trade-off between homophily (attribute-based choices) and degree-based attachment. This configuration exposes key behavioral differences while remaining computationally tractable for repeated simulations across models.

Before analyzing the assortativity patterns, we must clarify a change in the network size ($n$) and address the implications for the interpretation of our results. While the initial base case was conducted on networks of $n = 1500$, the extensive simulations involving attribute data were run on smaller networks of $n = 350$. The primary reason for this reduction was the substantial computational cost associated with running over one million API consultations.

This change in $n$ alters the relative size of the subgroup $s$. For $n = 1500$, a subgroup of $s = 50$ represents a small fraction ($\sim 3.3\%$) of the network, modeling limited information. For $n = 350$, $s = 50$ represents a much larger fraction ($\sim 14\%$), giving the joining node significantly higher visibility of the network's degree structure. However, this condition does not weaken our findings on homophily; on the contrary, it strengthens them. Our base case analysis shows that increased visibility (a larger $s$) amplifies the tendency for models to attach to the highest-degree nodes. By using $n = 350$ with $s = 50$, we create a more stringent test for attribute-based choices. In this scenario, the information about which nodes are the most ``popular'' is very clear and salient. For homophily to emerge, the LLM must actively ignore this strong structural signal in favor of attribute similarity. The fact that we observe powerful, statistically significant under these conditions demonstrates that the drive for homophily is a primary factor in the LLMs' decision-making, not just an artifact of limited information.

\subsection{Attribute and Degree Assortativity in LLM-Generated Networks}

In degree-exclusive scenarios, all models predominantly followed a highest-degree attachment strategy, where new nodes preferentially connected to the most well-connected existing nodes. However, when attributes are introduced, this pattern shifts. The models adopt more nuanced attachment behaviors, where connecting to the highest-degree nodes even those sharing similar attributes is no longer the sole priority. This indicates that attribute information interacts with structural preferences, leading to more diverse linking strategies in the generated networks. The LLMs begin to navigate a trade-off between degree and homophily, resulting in a set of complex, mixed attachment strategies.

\begin{figure}[h!]
    \centering
    \includegraphics[width=0.9\textwidth]{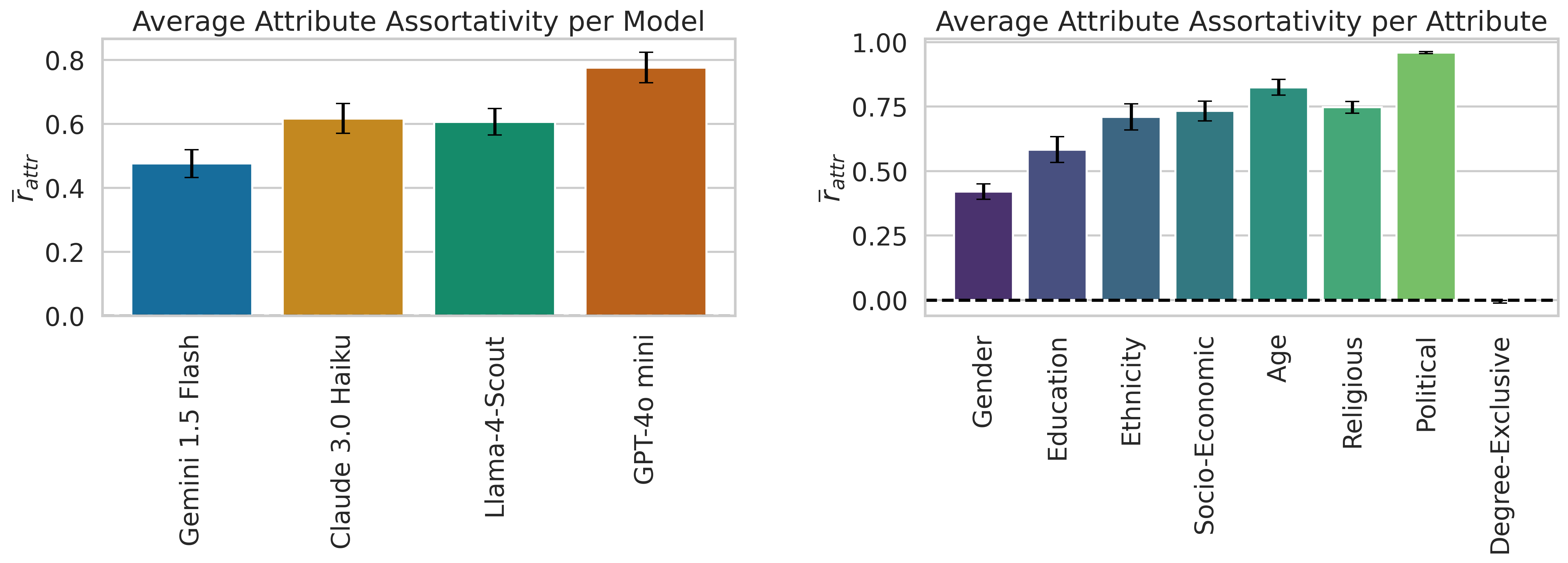}
    \caption{Average attribute assortativity (\(r_{\text{attr}}\)), the tendency of nodes to connect with others sharing similar attributes, in LLM-generated networks with \(n = 350\), \(m = 3\), \(s = 50\): (a) per model across Claude 3.0 Haiku, GPT-4o Mini, Gemini 1.5 Flash, and Llama-4-Scout; (b) per attribute across political orientation, gender, age, education, ethnicity, religious practice, and socio-economic status. Error bars denote standard error of the mean (SEM) and indicate variability in \(r_{\text{attr}}\) across attributes and models (right). A Welch's ANOVA indicated a highly significant main effect for model, \(F(3, 101.65) = 7.14\), \(p < .001\), and for category, \(F(7, 71.75) = 3073.58\), \(p < .001\).}
    \label{fig:assortativity_aggregates}
\end{figure}

\begin{figure}[h!]
    \centering
    \includegraphics[width=0.75\textwidth]{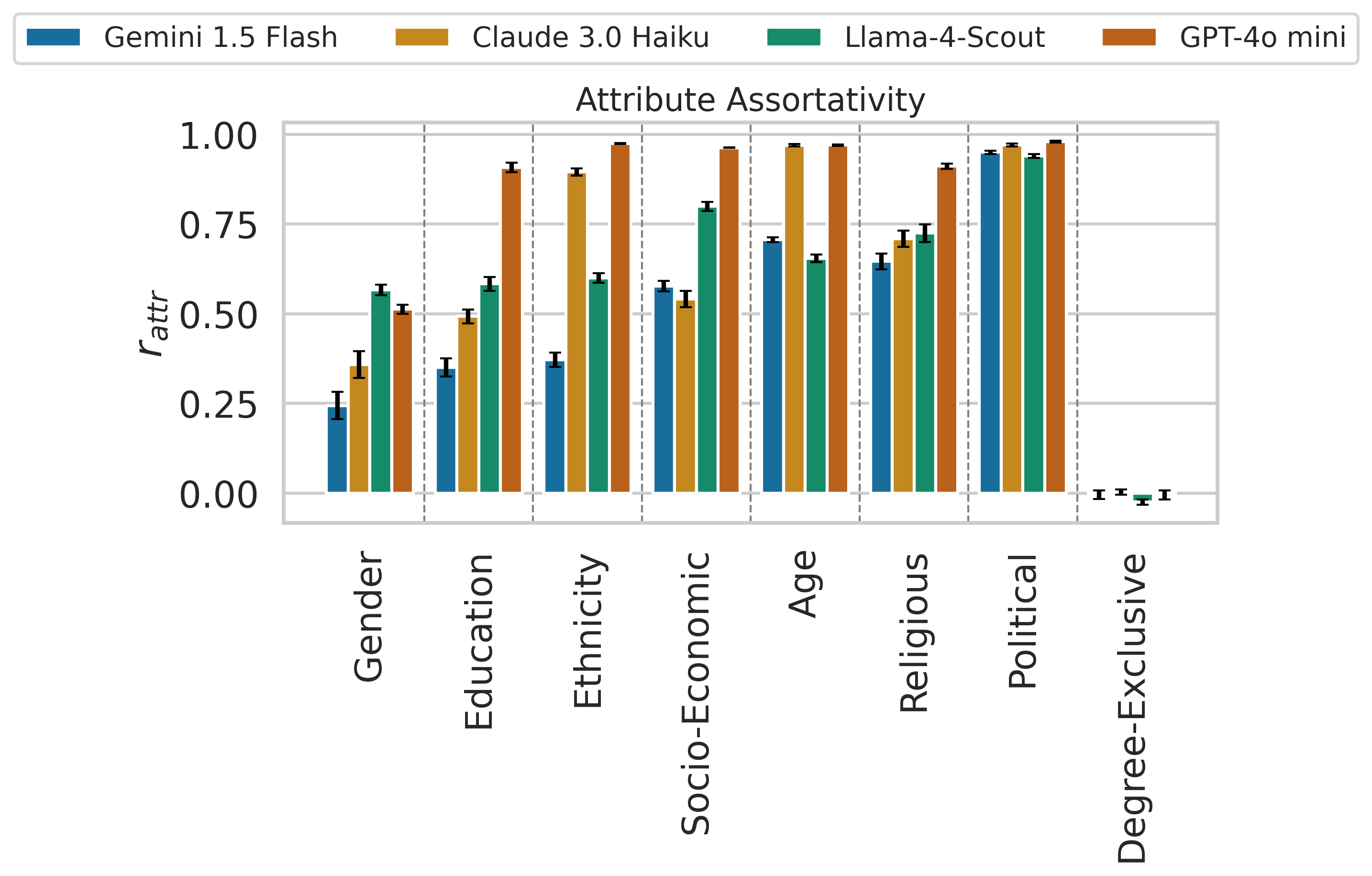}
    \caption{Attribute assortativity in LLM-generated networks with \( n = 350 \), \( m = 3 \), \( s = 50 \). This plot captures the extent to which nodes connect based on similarity in assigned node attributes. Error bars denote standard error of the mean (SEM)}
    \label{fig:attribute_assortativity}
\end{figure}

\begin{figure}[h!]
    \centering
    \includegraphics[width=0.75\textwidth]{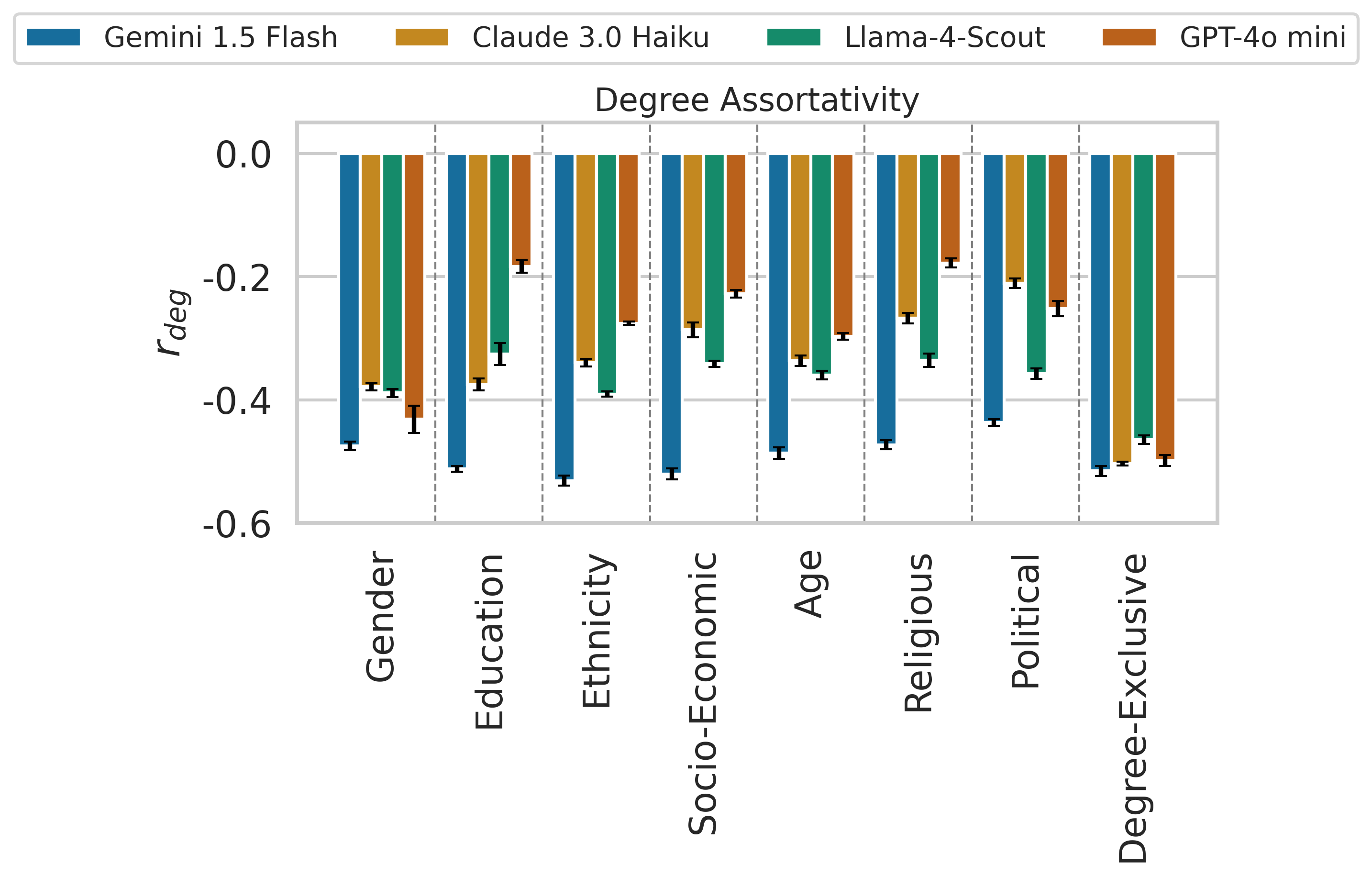}
    \caption{Degree assortativity in LLM-generated networks with \( m = 3 \), \( s = 50 \). This plot shows the tendency of nodes to connect to others with similar degree values. Error bars denote standard error of the mean (SEM).}
    \label{fig:degree_assortativity}
\end{figure}

We examine the assortativity patterns in networks generated by large language models, specifically focusing on degree assortativity \(r_{\text{deg}}\), the tendency of nodes to connect with others having a similar number of connections, and attribute assortativity \(r_{\text{attr}}\), the tendency of nodes to connect with others having a same attribute, across four models: Gemini 1.5 Flash, GPT-4o Mini, Claude 3.0 Haiku, and Llama-4-Scout. Figure~\ref{fig:assortativity_aggregates} presents the average \(r_{\text{attr}}\) per model and per attribute, providing insights into the overall homophily tendencies in LLM-generated networks under the configuration \(m = 3, s = 50\). Additionally, Figure~\ref{fig:attribute_assortativity} and Figure~\ref{fig:degree_assortativity} present a comparative assessment of assortativity across multiple attribute dimensions. For the exact values of degree and attribute assortativity across all models and attributes, please refer to Table~\ref{tab:assortativity_summary} in the appendix.

Figure\hyperref[fig:assortativity_aggregates]{~\ref*{fig:assortativity_aggregates}(a)} shows that different LLMs create networks with different degree of attribute-based homophily. GPT-4o Mini exhibits the highest average \(r_{\text{attr}}\) , indicating a strong tendency for nodes with similar attributes to connect, reflecting pronounced homophily. Claude 3.0 Haiku and Llama-4-Scout indicating a weaker but still positive homophilous tendency. Gemini 1.5 Flash displays the lowest average \(r_{\text{attr}}\), suggesting the least homophilous behavior among the models, though still positive, indicating some preference for same-attribute connections. The distinct separation between models highlights fundamental differences in how each LLM prioritizes attribute-based mixing.

Figure\hyperref[fig:assortativity_aggregates]{~\ref*{fig:assortativity_aggregates}(b)} presents the average \(r_{\text{attr}}\) aggregated across all models for each attribute, revealing the relative influence of different attributes on homophily. Education and Gender display the lowest average \(r_{\text{attr}}\), indicating weaker homophilous tendencies for these attributes. The strongest drivers of homophilous connections across all models is Political Orientation, exhibiting the highest average \(r_{\text{attr}}\), which is one of our two value attributes, indicating that LLMs' attention mechanism pays more attention to commonalities in less static shared beliefs and attitudes than in static shared social characteristics. There is not statistically significant difference among Age, Ethnicity, Socio-Economic status, and our other value-based attribute, namely Religious Practice. However, all of them  influence assortative mixing. 

When comparing the choices of the different LLMs as can be seen in Figure~\ref{fig:attribute_assortativity}, GPT-4o Mini consistently demonstrates the highest attribute assortativity, with values approaching \(1.0\) for almost all of the attributes reflecting strong homophilous tendencies. Conversely, the previously detected low value for Gemini 1.5 Flash arises from more variability across attributes, such as a low \(r_{\text{attr}}\) for gender and a high value for political orientation. 

\begin{figure}[h!]
    \centering
    \includegraphics[width=0.9\textwidth]{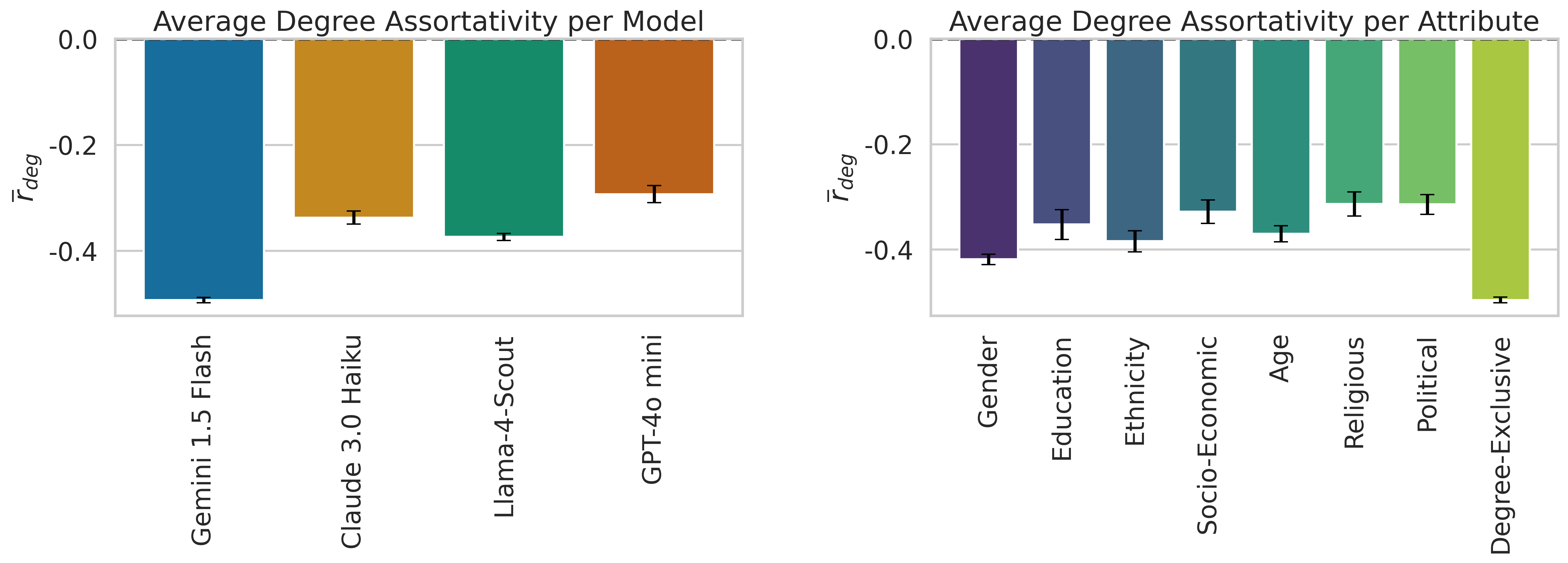}
    \caption{Average degree assortativity (\(r_{\text{deg}}\)) in LLM-generated networks with \(n = 350\), \(m = 3\), \(s = 50\): (a) per model across Claude 3.0 Haiku, GPT-4o Mini, Gemini 1.5 Flash, and Llama-4-Scout; (b) per attribute across political orientation, gender, age, education, ethnicity, religious practice, and socio-economic status. Error bars denote standard error of the mean (SEM) and indicate variability in \(r_{\text{deg}}\) across attributes (a) and models (b). A Welch's ANOVA indicated a highly significant main effect for model, F(3, 95.99) = 116.17, p < .001, and for category, F(7, 72.84) = 34.71, p < .001.}
    \label{fig:degree_assortativity_aggregates}
\end{figure}

Figure\hyperref[fig:degree_assortativity_aggregates]{~\ref*{fig:degree_assortativity_aggregates}(a)} reveals that LLMs differ in how they integrate degree information when forming links, when attributes are provided. We find consistent negative average degree assortativity \( r_{\text{deg}} \) across all tested LLMs, indicating a general tendency for nodes of dissimilar degrees to connect in generated networks even when attributes are provided. Among the models, Gemini 1.5 Flash yields the most disassortative structures, while GPT-4o Mini shows the weakest disassortative tendencies. The lowest values are observed for gender and ethnicity, implying that these attributes are more likely to lead to degree-disassortative connections, see Figure\hyperref[fig:degree_assortativity_aggregates]{~\ref*{fig:degree_assortativity_aggregates}(b)}. This may indicate that LLMs rely less on degree when evaluating connections involving highly salient or demographically distinct categories. 

In a more detailed picture that displays per model per attribute behaviours, the degree assortativity \(r_{\text{deg}}\), as shown in Figure~\ref{fig:degree_assortativity}, consistently takes negative values across all models and attributes. This reflects a disassortative mixing pattern, in which low-degree (incoming) nodes tend to connect with high-degree nodes. Among the models, Gemini 1.5 Flash exhibits the strongest disassortativity. In contrast, GPT-4o Mini shows the least negative \(r_{\text{deg}}\), indicating a weaker disassortative tendency.

For degree assortativity, attributes like gender, age, ethnicity exhibit the strongest disassortative behavior across all models. In contrast, attributes such as political orientation and religious practice show less negative \(r_{\text{deg}}\), suggesting that these attributes may have a weaker influence on degree-based mixing patterns.

\subsubsection{Statistical Significance}

This section presents the results of pairwise independent samples \textit{t}-tests comparing the average degree assortativity coefficients and average attribute assortativity coefficients across different models and attribute categories. Welch’s \textit{t}-test was used in all comparisons, as it does not assume equal variances between groups and is more robust in the presence of heteroscedasticity~\cite{welch1947generalization}. These tests allow us to identify statistically significant differences in assortativity patterns, both in terms of structural connectivity and node attributes. To account for the increased risk of errors when conducting multiple simultaneous tests, we applied the Holm-Bonferroni correction to the set of p-values from our pairwise comparisons. This method controls the family-wise error rate.

\begin{figure}[h!]
    \centering
    \begin{subfigure}[b]{0.48\textwidth}
        \centering
        \includegraphics[width=\textwidth]{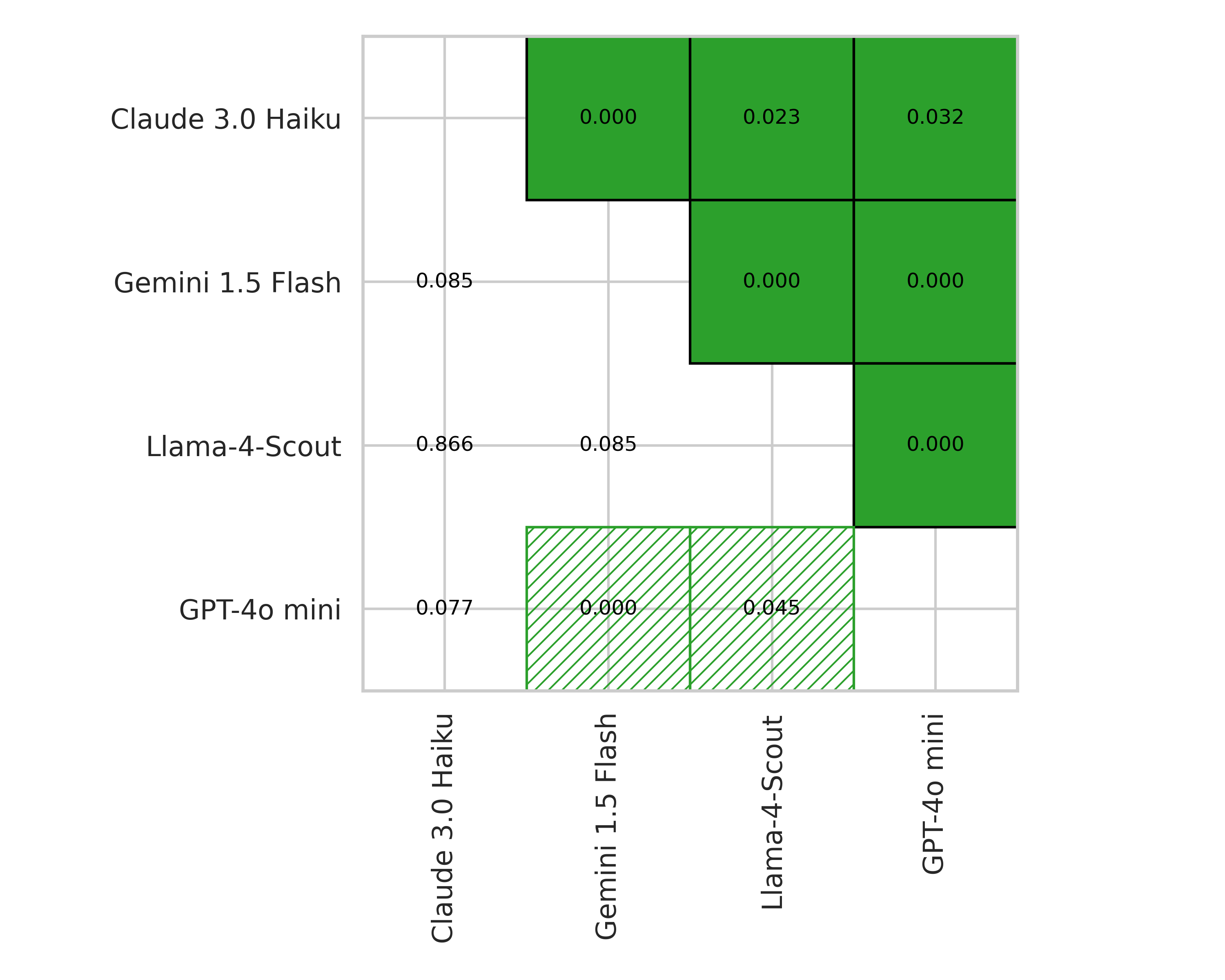} 
        \caption{Pairwise comparisons between LLM models}
        \label{fig:model_pairwise_combined_heatmap}
    \end{subfigure}
    \hspace{-0.05\textwidth}
    \begin{subfigure}[b]{0.48\textwidth}
        \centering
        \includegraphics[width=\textwidth]{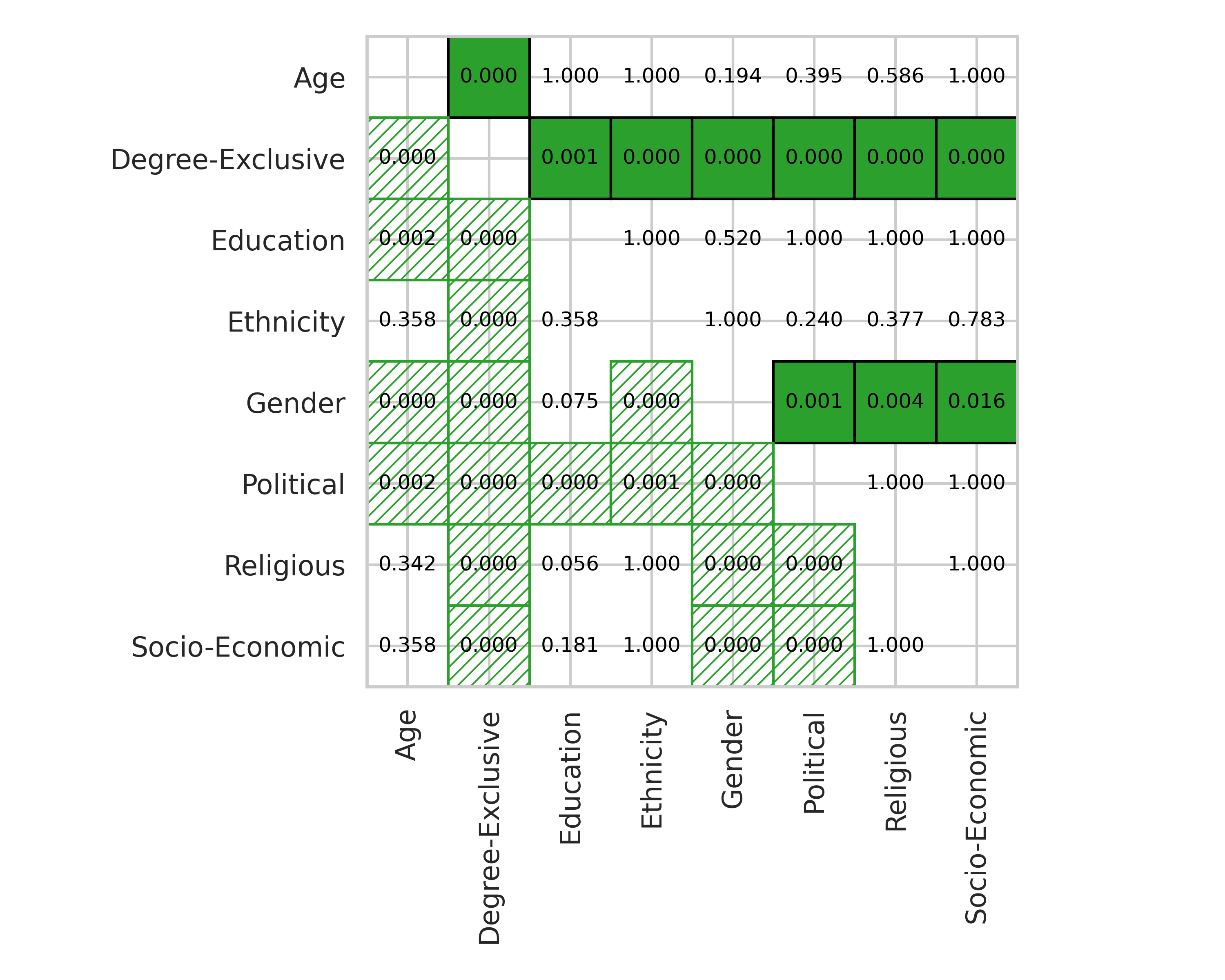} 
        \caption{Pairwise comparisons between attribute categories}
        \label{fig:category_pairwise_combined_heatmap}
    \end{subfigure}
    \caption{
    Pairwise Welch’s \textit{t}-test results for network assortativity metrics. Reported values are Holm-Bonferroni corrected \textit{p}-values from statistical comparisons. Statistically significant results after correction are highlighted in green. The \textit{upper triangle}"(solid fill)" represents pairwise comparisons of average \textit{degree assortativity}, while the \textit{lower triangle}"(hatched fill)" corresponds to average \textit{attribute assortativity}, where (\textit{df} = 91) and (\textit{df} = 35), respectively. A low adjusted p-value indicates strong evidence to reject the null hypothesis of equal group means, while a high p-value indicates insufficient evidence to reject it.
    }
\label{fig:overall_pairwise_combined_assortativity_heatmaps}
\end{figure}

As shown in Figure~\ref{fig:overall_pairwise_combined_assortativity_heatmaps}, there are numerous statistically significant pairwise differences in both attribute and degree assortativity. Notably, all model comparisons are statistically significant, except for the attribute assortativity between \textit{LLaMA}-\textit{Claude}, \textit{LLaMA}-\textit{Gemini},  \textit{Claude}-\textit{Gemini} and \textit{GPT-4o}-\textit{Claude}, which did not reach significance. Among attribute categories, the majority of pairwise comparisons show significant differences in attribute assortativity, with exceptions such as  \textit{Ethnicity–Age}, \textit{Ethnicity–Education}, \textit{Ethnicity–Religious}, \textit{Socioeconomic–Age}, \textit{Socioeconomic–Ethnicity}, and \textit{Socioeconomic–Religious}. In contrast, degree assortativity comparisons across attribute categories reveal a mixture of significant and non-significant results, suggesting more heterogeneous structural differentiation in terms of node connectivity patterns.

\subsection{Tie Formation Scheme}

This section investigates tie formation rates in networks generated by large language models. The analysis focuses on the relative frequencies of connections between nodes categorized by binary attributes. The tie formation rates are categorized into homophilous types and heterophilous type, each in two varieties (following the definitions from Table~\ref{table:binary_attributes}), resulting in four types: 
\begin{itemize}
    \item two homophilous: cat 1 $\to$ cat 1 (e.g., liberal $\to$ liberal), and cat 2 $\to$ cat 2 (e.g., conservative $\to$ conservative), and
    \item two heterophilus: cat 1 $\to$ cat 2 (e.g., liberal $\to$ conservative), cat 2 $\to$ cat 1 (e.g., conservative $\to$ liberal).
\end{itemize}

\begin{figure}[h!]
    \centering
    \includegraphics[width=0.9\textwidth]{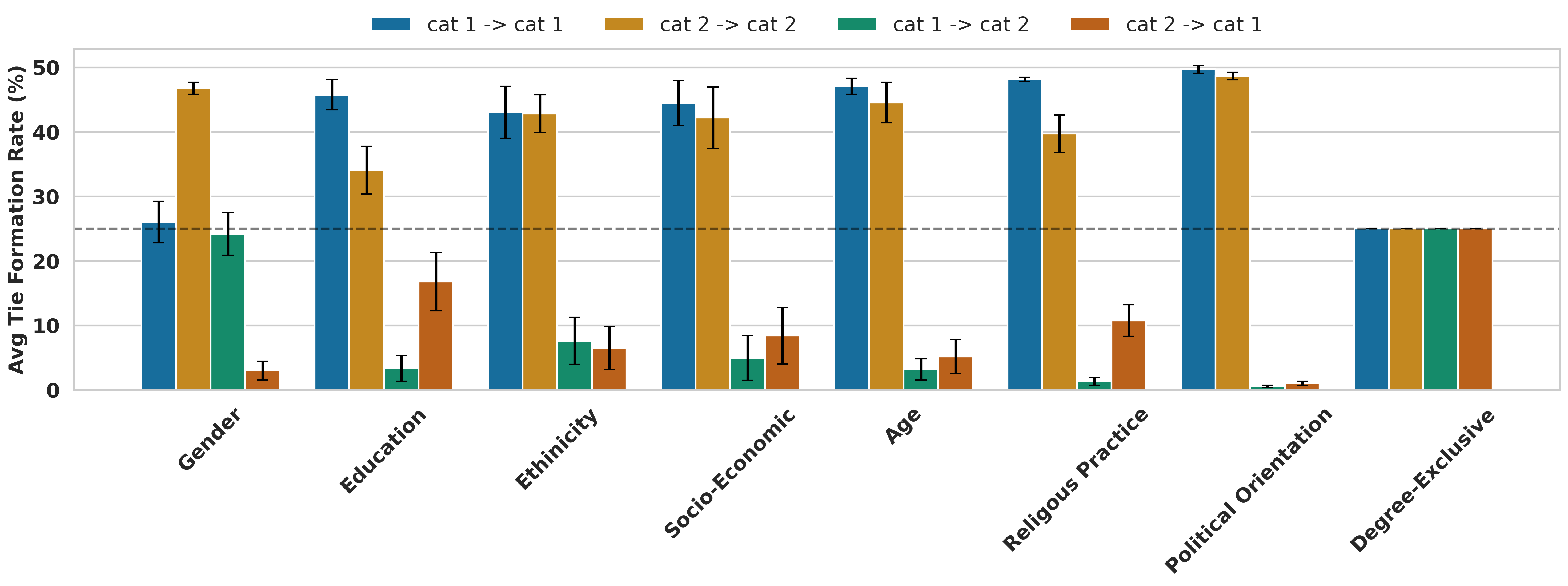}
\caption{Average tie formation rates in LLM-generated networks for various binary attributes: political orientation, gender, age, education, ethnicity, religious practice, and socio-economic status. A synthetic "Degree-Exclusive" attribute is also included, representing a neutral baseline where all tie formation rates are fixed at 0.25. The rates are averaged across four models (Gemini 1.5 Flash, GPT-4o Mini, Claude 3.0 Haiku, Llama-4-Scout) and categorized by tie directions. Error bars denote standard error of the mean (SEM). Category definitions for each attribute are provided in Table~\ref{table:binary_attributes}.}
    \label{fig:tie_formation_avg_across_models}
\end{figure}

In addition to the summary picture that attribute assortativity provides, examining directional tie formation patterns allows us to uncover potential asymmetries in homophily and cross-group interactions, revealing potential biases in how different LLMs simulate social connection dynamics. The rates, averaged over the models, see Figure~\ref{fig:tie_formation_avg_across_models}, reveal a pronounced homophilous tendency, with cat 1 $\to$ cat 1 and cat 2 $\to$ cat 2 connections consistently dominating at approximately 40\% to 50\% across most attributes, while heterophilous ties (cat 1 $\to$ cat 2 and cat 2 $\to$ cat 1) remain low. The consistently low heterophilous rates across all attributes might indicate a limited capacity for LLMs to model diverse, cross-category interactions under this configuration. 

It should be noted that the network graph itself is undirected. When a new node establishes a connection, a single, undirected edge is added to the graph. The structural properties analyzed in Section 4 such as the clustering coefficient, diameter, and average shortest path length are therefore calculated for an undirected graph. However, the analysis of 'tie formation rates' in this section leverages directional information inherent to the network's growth process. At each time step, a new 'source' node initiates a connection with an existing 'target' node. By tracking the attributes of both the source and the target of this decision, we can analyze asymmetries in the generative mechanism itself.

The first finding in distinguishing between formation schemes is the revelation that the high attribute assortativity of Political Orientation (see Figure\hyperref[fig:assortativity_aggregates]{~\ref*{fig:assortativity_aggregates}(b)}), seems to stem from a pronounced difference between homophilous ties (very high) and heterophilous ties (very low). There are many assortative ties, but very few disassortative mixing when LLMs create networks of political orientation. Figure~\ref{fig:tie_formation_avg_across_models} also reveals a subtle asymmetry in tie formation rates for the gender attribute, with the male $\to$ male rate at approximately 25\%, compared to the female $\to$ female rate of around 45\%. This disparity indicates stronger homophilous tendencies among female nodes than male nodes. A similar, though less pronounced, pattern is observed for the education and religious practice attributes. Nodes with a college degree exhibit a higher propensity to form connections with similarly educated peers compared to those without a degree. Likewise, in religious practice, non-practicing nodes show a preference for homophilous ties (non-practicing $\to$ non-practicing), whereas practicing nodes are more active in forming ties with other practicing nodes (practicing $\to$ practicing).

Another notable finding is the near-symmetry in homophilous tie formation rates among the most attributes. Across most models and attributes, the cat 1 $\to$ cat 1 and cat 2 $\to$ cat 2 rates are remarkably balanced. This symmetry suggests that LLMs do not exhibit a bias toward one category over the other within an attribute, treating both categories as equally likely to form homophilous connections. This balance is promising for applications where fairness and impartiality in network generation are critical, such as in social simulation or recommendation systems. 

Maybe the most interesting finding is the conclusion that The variety arises from differences in how LLMs form heterophilous connections (cat 1 $\to$ cat 2 and cat 2 $\to$ cat 1). For example, for gender, education, and religious practices, one disassortative tie dominates the other. In specific:

\begin{itemize}
    \item \textit{Gender:} The rate of "male $\to$ female" tie formation is significantly higher than "female $\to$ male". This implies that individuals male are more likely to form connections with individuals in female than vice versa.
    \item \textit{Education:} "no-degree $\to$ college-degree" tie formation is also higher than "college-degree $\to$ no-degree". This mean that individuals without a college-degree are more prone to connect with those with a college-degree compared to the reverse.
    \item \textit{Religious Practice:} Similar to gender and education, "non-practicing $\to$ practicing" ties are more prevalent than "practicing $\to$ non-practicing" ties. This suggests a directional preference in tie formation based on religious practices, where non-practicing individuals initiates more connections with the practicing compared to the reverse.
\end{itemize}

These asymmetries in tie formation have direct structural consequences. A supplementary analysis, detailed in Appendix~\ref{app:attribute_degree_correlation}, confirms that the categories acting as preferential targets for connections accumulate a statistically significant higher average degree. For example, the greater prevalence of  "male $\to$ female" connections results in female nodes achieving a higher final centrality in the network. This provides a direct, quantitative link between the micro-level connection biases of the LLMs and the macro-level structural inequalities that emerge.

Reflecting on the meanings of these identified asymmetries, each one makes sense, as the direction goes to what is usually considered to be the 'more attractive' direction (figuratively speaking, from a social semantic perspective). It is interesting to discover that the 'micromotives' of the individual choices of LLMs result in such coherent 'macrobehavior'~\cite{schelling2006micromotives} when looked at from this larger scale. From this larger perspective we can conclude that, confronted with a binary choice in these areas of the social fabric, LLMs see the largest attractor pull exercised by women attracting men, and by those with a college-degree attracting the connection from those without (see Fig.~\ref{fig:tie_formation_avg_across_models}). See Figure~\ref{fig:connection_distribution_grid} in the appendix for a detailed analysis of tie formation across all the different models and attributes.

The consistent directional bias in heterophilous connections where, for example, males are more likely to initiate ties with females, and the non-college-degreed with the college-degreed has profound implications for the large-scale AI-driven networks. If such asymmetries scale, they are likely to produce and entrench significant structural inequalities. Over time, the groups that are the preferential target of incoming ties would accumulate disproportionately high centrality. This could create a stable core-periphery structure, where socially 'attractive' groups (e.g., professionals, the college-educated) form a central, highly connected core that enjoys privileged access to information and influence, while other groups remain on the periphery.

Furthermore, the interaction between these biased attributes could create even starker stratification. For instance, our findings indicate that both female nodes and professional nodes act as strong attractors for ties. A critical question for future work is how these effects combine: would a 'professional female' node become a super-attractor, concentrating network influence at the intersection of these two attributes? Such combinatorial effects could amplify inequality beyond what is observed with a single attribute, leading to a network where a small subset of nodes at the intersection of 'desirable' categories holds the vast majority of social capital. These emergent hierarchies, born from subtle, directional micro-preferences embedded in the LLMs, could systematically reinforce and even exacerbate existing societal inequalities in digital ecosystems governed by AI agents.

\section{Structural Properties of Generated Networks}

\begin{figure}[h!]
    \centering
    \begin{subfigure}[t]{0.48\textwidth}
        \centering
        \includegraphics[width=\textwidth]{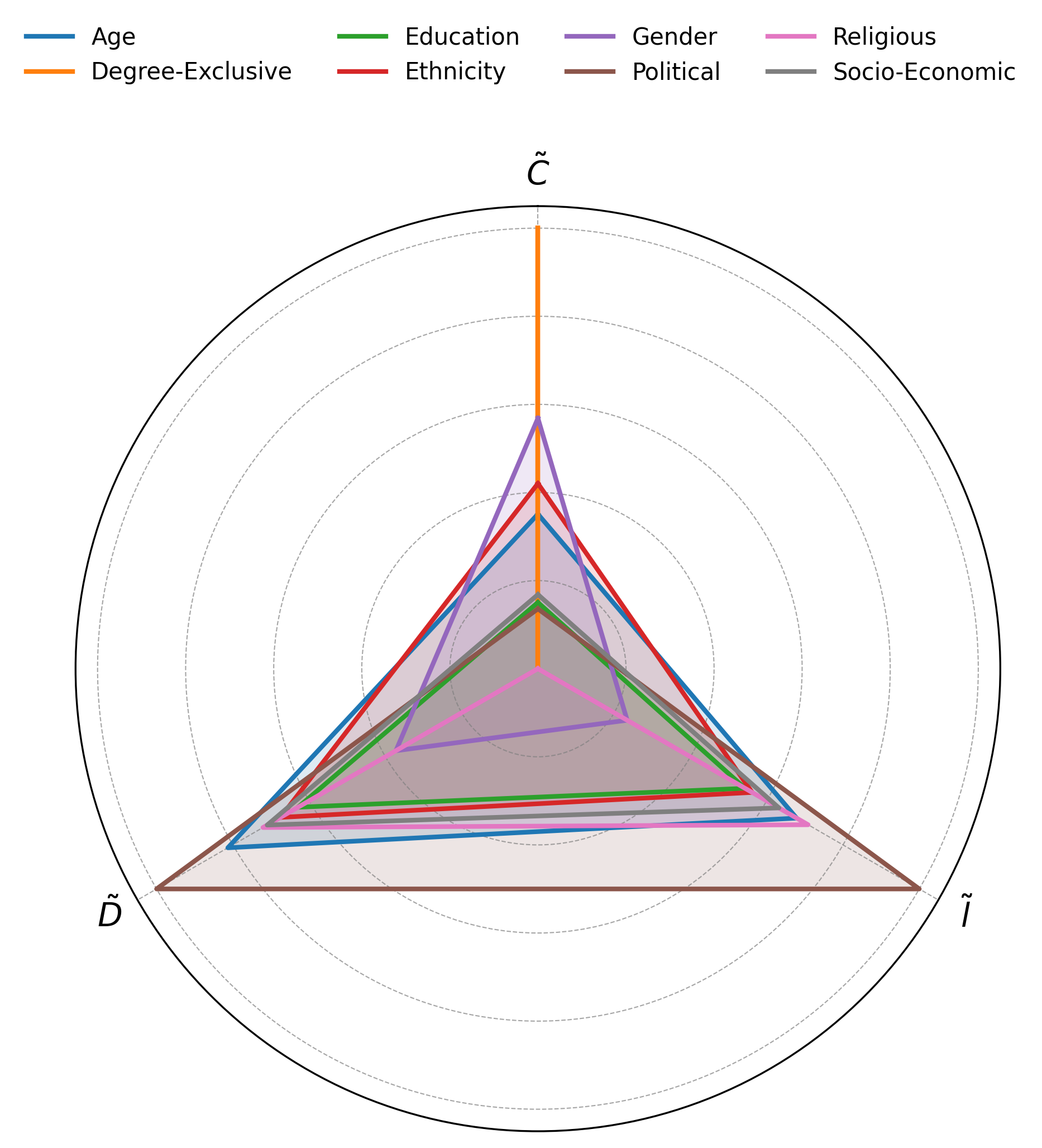}
        \caption{Radar plot of normalized metrics: average clustering coefficient (\(\tilde{C}\)), average shortest path length (\(\tilde{l}\)), and diameter (\(\tilde{D}\)).}
        \label{fig:radar-metrics}
    \end{subfigure}
    \hfill
    \begin{subfigure}[t]{0.48\textwidth}
        \centering
        \includegraphics[width=\textwidth]{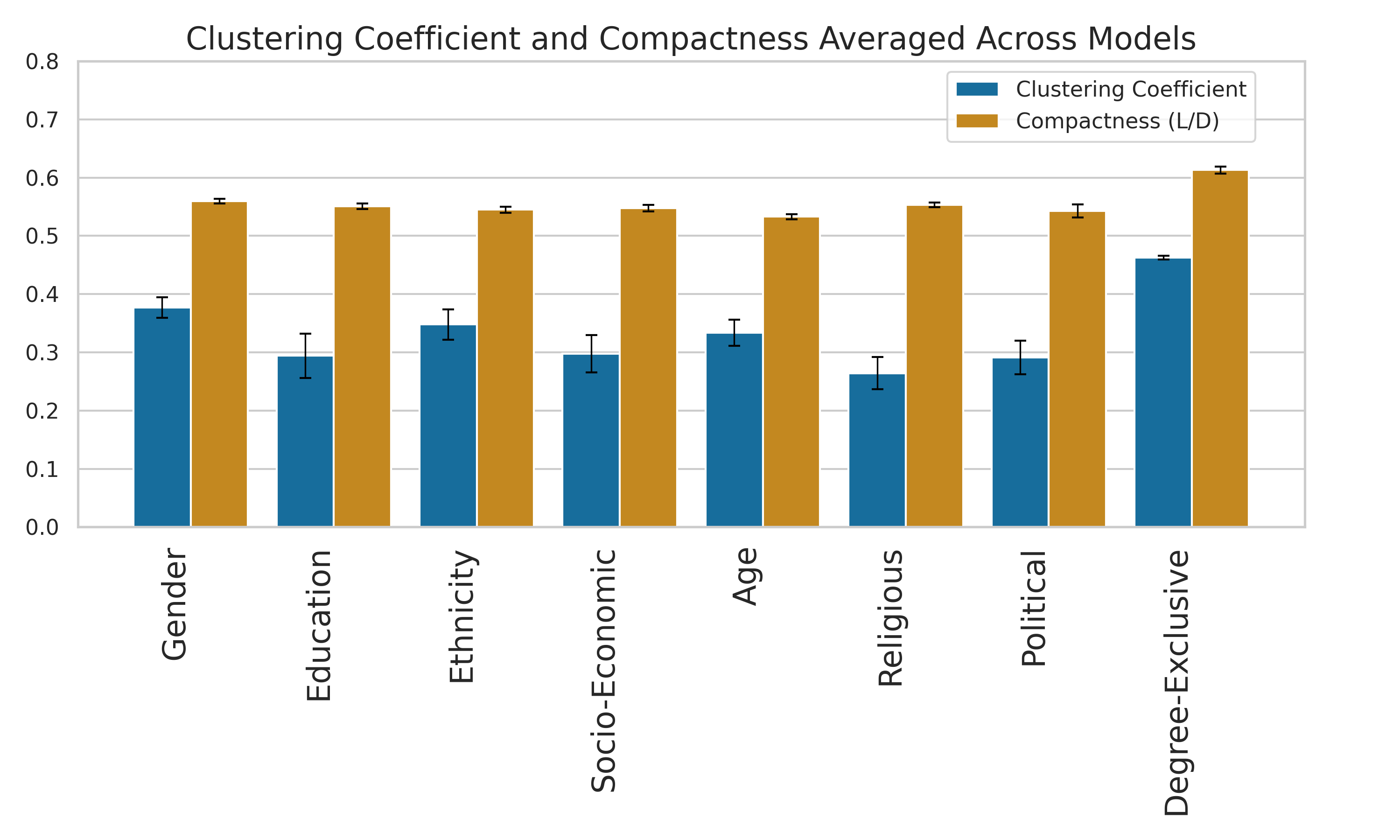}
        \caption{Bar plot of average clustering and compactness per attribute across all models.}
        \label{fig:compactness-bar}
    \end{subfigure}
    \caption{Comparison of structural metrics in LLM-generated networks with \(n = 350\), \(m = 3\), \(s = 50\). Error bars denote standard error of the mean (SEM).}
    \label{fig:network-structure-comparison}
\end{figure}

This section evaluates the structural properties of networks generated by large language models (LLMs). The average clustering coefficient measures the tendency of nodes to form tightly knit groups. Specifically, we examine three key metrics: the average clustering coefficient, which measures the tendency of nodes to form tightly knit groups; the average shortest path length, which captures the typical distance between pairs of nodes; and the diameter, which represents the longest shortest path within the network. The plot in Figure~\ref{fig:radar-metrics} visualizes the normalized values (via min-max scaling) of three network metrics: normalized average clustering coefficient (\(\tilde{C}\)), normalized average shortest path length (\(\tilde{l}\)), and normalized diameter (\(\tilde{D}\)). Among these, the clustering coefficient (\(\tilde{C}\)) is considered to be a so-called local metric, because it is fundamentally based on local neighborhood structures, measureing how tightly connected each node's neighbors are to one another, and, hence, capturing local properties around individual nodes. The shortest path length (\(\tilde{l}\)), and normalized diameter (\(\tilde{D}\)) are considered to be global metrics, because their calculation requires information about the entire network. We aggregate these metrics per attribute, providing a comparative view of structural properties across all models in LLM-generated networks.

We observe that the \textit{Gender} exhibits the highest normalized average clustering coefficient (\(\tilde{C}\)) after the degree-exclusive, indicating a strong tendency for nodes within this attribute to form tightly knit clusters. In contrast, the \textit{Religious Practice} shows the lowest (\(\tilde{C}\)), implying weaker clustering tendencies and potentially more fragmented structures. Regarding the normalized average shortest path length (\(\tilde{l}\)), the \textit{Political Orientation} stands out with the highest value, suggesting that paths between nodes in this attribute are relatively longer, which may indicate a more sparse or elongated network structure. On the other hand, the \textit{Gender} has the lowest \(\tilde{l}\). For the normalized diameter (\(\tilde{D}\)), we see similar behaviour. 

The bar plot in Figure~\ref{fig:compactness-bar} illustrates the average compactness (\(L/D\)) and clustering coefficient across various attributes in LLM-generated networks. We observe that, despite variations in the diameter (\(D\)) and average shortest path length (\(L\)), the compactness remains relatively consistent across all attributes. In contrast, the clustering coefficient exhibits more pronounced variations across the attributes. This suggests that the inclusion of different attributes may lead LLMs to adopt distinct attachment decision-making strategies, which, while maintaining similar levels of compactness, result in significantly different clustering behaviors. These findings indicate that attribute-specific influences in LLMs may disproportionately affect local network structures (e.g., clustering) while preserving global structural properties (e.g., compactness), meriting further exploration into the underlying generative processes. 

\begin{figure}[h!]
    \centering
    \includegraphics[width=0.9\textwidth]{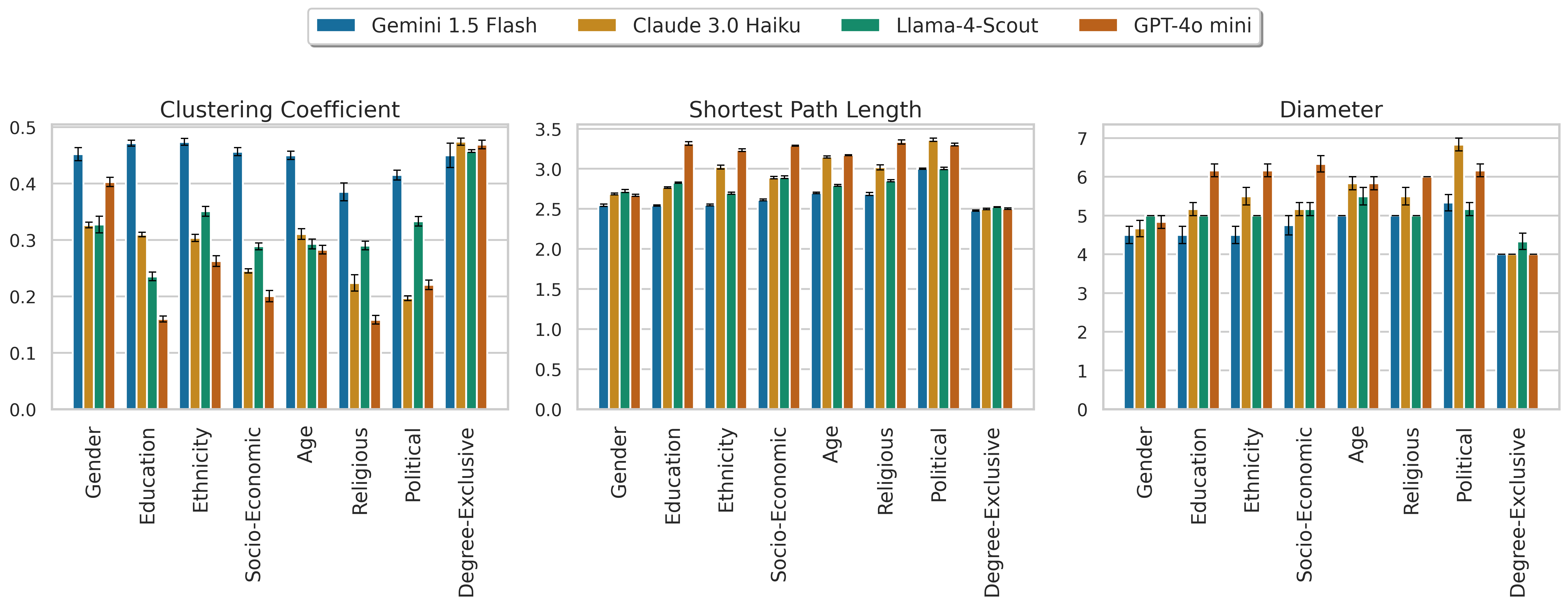}
    \caption{Comparison of network structure metrics across four language models (Gemini 1.5 Flash, GPT-4o mini, Claude 3.0 Haiku, and Llama-4-Scout) for different node attributes with \(n = 350\), \(m = 3\), \(s = 50\). Each subplot shows a different metric: (left) average clustering coefficient, (center) average shortest path length, and (right) average diameter. Error bars denote standard error of the mean (SEM).}
    \label{fig:network-metrics}
\end{figure}

Table~\ref{tab:network-metrics-summary} provides a detailed breakdown of the network metrics for each model across various attributes, complementing the visual representation in Figure~\ref{fig:network-metrics}. This per-model, per-attribute analysis reveals distinct structural characteristics and variations in the networks generated by each LLM. Across all models, the \textit{Degree-Exclusive} scenario consistently exhibits the highest average clustering coefficient. This indicates that when attachment decisions are based solely on degree, the resulting network structure is characterized by highly cohesive clusters, likely due to strong preferential attachment mechanisms in the generative process. Conversely, attributes like \textit{Religious} and \textit{Education} tend to show lower clustering coefficients suggesting more fragmented local structures for these attributes in this model.

For the average shortest path length, Gemini 1.5 Flash generally produces the most compact networks. In contrast, GPT-4o mini and Claude 3.0 Haiku tend to have longer average shortest path lengths, particularly for the \textit{Political} attribute indicating more elongated network structures and potentially lower connectivity. The average diameter also shows notable variation. GPT-4o mini and Claude 3.0 Haiku generate networks with larger diameters for the \textit{Political} attribute, suggesting greater structural dispersion and a larger extent of the network. On the other hand, the \textit{Degree-Exclusive} attribute consistently has the smallest diameter across all models, reinforcing the compact and centralized nature of networks without any attribute.

Comparing the models, Gemini 1.5 Flash stands out for producing networks with higher clustering coefficients and shorter average shortest path lengths across most attributes, indicating a tendency to generate more compact and clustered structures. Llama-4-Scout, while similar in some respects, shows more moderate clustering and slightly larger diameters, suggesting a balance between compactness and dispersion. GPT-4o mini and Claude 3.0 Haiku, however, exhibit greater variability across attributes, with GPT-4o mini achieving the highest clustering for \textit{Gender}, but also producing more dispersed networks for attributes like \textit{Socio-Economic}.

\section{Conclusion} 

This study reveals a novel understanding of how multi-agent systems powered by large language models (LLMs) fundamentally reshape network structures through emergent, socially-biased behaviors embedded in their vast parameter spaces. We demonstrate that these LLM-driven agents do not simply replicate analytically derived network dynamics from the textbook, nor do they merely produce similar networks when prompted identically; instead, they create significantly different network structures. We found that different AIs create significantly different network structures and that when they evaluate different societal aspects, they give them significantly different weights. 

We started by showing that in the base-case, LLM-driven agents follow the theoretically expected case of inherently favor high-degree nodes, reflecting a strong form of preferential attachment. However, when social attributes such as age, gender, religion, and political orientation are introduced, network evolution deviates significantly from baseline topology in distinct and insightful ways. This will increasingly make a notable difference in the social fabric as AI agents are increasingly being tasks with taking autonomous network decisions.

One core finding is that these attributes induce homophilic clustering and heightened assortativity. The LLMs' decision-making shifts from a simple heuristic to a set of nuanced, mixed strategies that balance degree and social similarity, with political and religious dimensions leading to the most pronounced fragmentation and polarization. Age and gender, by contrast, generate more gradual and diffuse structural shifts. It is important that this is a choice that arises from the LLMs training data and fine-tuning, and not inevitable. Notably, we find a distinction in the effects of different types of attributes, where 'value homophily' (e.g., political and religious beliefs) acts as a much stronger driver of network fragmentation than 'status homophily' (e.g., age and gender). This suggests that LLMs, in simulating human sociality, prioritize shared ideologies over shared demographic characteristics when forming connections.

Across models, we observe that generative AI agents do not act as neutral conduits of connectivity; instead, they embed and amplify human-like biases, creating self-organized patterns that mirror social divisions. Political science studies have long documented stark homophily and polarization in political affiliations, especially in online and social network environments, contrasting with more moderate or context-dependent homophily in attributes like gender or age. For instance, one study analyzed exposure to ideologically diverse news on Facebook and found that users tended to interact more with like-minded individuals and content, especially in political contexts~\cite{bakshy2015exposure}. Similarly, another study on  Twitter retweet and mention networks during the 2010 U.S. midterm elections and found strong ideological clustering such that retweets overwhelmingly occurred within political communities rather than across them, revealing high political homophily and echo chambers~\cite{conover2011political}. By contrast, research on gender-based homophily suggests more nuanced patterns. While some degree of gender homophily exists such as in friendship networks~\cite{mcpherson2001birds} it is often less rigid and varies with context, such as workplace structure or domain-specific settings. In mixed-gender environments, for example, collaborative or professional goals often dilute homophily effects~\cite{ibarra1992homophily}. Thus, we observe a certain degree of resonance between social networks and LLM-generated networks.

Our analysis further reveals substantial differences in the network structures generated by various LLMs. Our findings are complementary to recent research by De Marzo et al.~\cite{de2023emergence}, who demonstrated that LLMs can replicate the mechanism of preferential attachment to form scale-free networks, provided that technical biases like token priors are controlled for. While their work confirms the LLM's capacity for degree-based social dynamics, our study shows that introducing socially meaningful attributes causes a significant deviation from these baseline models. We find that LLMs embed and enact powerful homophilic preferences, creating networks defined by social clustering and asymmetric ties that reflect societal norms. Taken together, these studies suggest that LLM-based multi-agent systems are not monolithic; their emergent network structures are a complex product of technical artifacts, fundamental network growth principles, and deeply embedded social biases.

Among the models evaluated, \texttt{GPT-4o Mini} consistently exhibits the highest average attribute assortativity, indicating a strong propensity for homophily where nodes with similar attributes preferentially connect. This model frequently yields assortativity values approaching 1.0 across a broad range of attributes. In comparison, \texttt{Claude 3.0 Haiku} and \texttt{Llama-4-Scout} demonstrate moderate yet consistently positive assortativity values, positioning them between \texttt{GPT-4o Mini} and \texttt{Gemini 1.5 Flash}. \texttt{Gemini 1.5 Flash} shows the lowest average attribute assortativity, suggesting weaker homophilous tendencies, though its assortativity values are more variable across different attributes. Pairwise Welch’s \textit{t}-tests confirm that these differences in both attribute and degree assortativity are statistically significant across most model comparisons, with the exception of attribute assortativity between \texttt{Claude 3.0 Haiku} and \texttt{Llama-4-Scout}. 

In addition to the overarching patterns of homophily, our study reveals that LLM-generated networks exhibit striking asymmetries in tie formation. For attributes like Gender, Education, and Religious Practice, the connection rates across categories are not balanced. For example, connections from male agents to female agents are far more prevalent than the reverse. This directional preference in heterophilous ties necessarily creates a corresponding asymmetry in the strength of in-group homophily. The critical finding is this consistent, directional bias in how LLMs form connections across social categories. Specifically, attributes such as \textit{Gender}, \textit{Education}, and \textit{Religious Practice} demonstrate directional biases, where ties from Category 2 to Category 1 (e.g., male~$\rightarrow$~female, non-college~$\rightarrow$~college) are more prevalent than the reverse. This suggests that the LLMs do not treat disassortative connections as symmetric interactions, but instead encode or reproduce directional preferences. In terms of structural properties, \texttt{Gemini 1.5 Flash} tends to generate more compact and highly clustered networks, as reflected by higher clustering coefficients and shorter average shortest path lengths. Conversely, \texttt{GPT-4o Mini} and \texttt{Claude 3.0 Haiku} often produce more diffuse networks, characterized by longer average shortest paths and greater diameters, particularly for attributes such as \textit{Political Orientation}.

Another core finding is that attribute-specific influences in LLMs seem to disproportionately affect local network structures (e.g., clustering) while preserving some global structural properties (e.g., compactness). This observation is particularly salient given that our model does not explicitly incorporate distance metrics in the network generation process, and we rely on partial, rather than global full, network information during generation. There could be several reasons for this asymmetric impact. The partial information available to the LLM might  naturally lead to a focus on immediate neighbors or hubs and their attributes, reinforcing these local connections. Moreover, global structural features such as average shortest path length or network diameter are often emergent properties of broader generative dynamics. These dynamics, which might involve a complex interplay of various factors beyond explicit attribute alignment, may be less sensitive to the specific attribute biases captured during node generation. The absence of explicit distance calculations could also mean that the LLM's generative process, while shaping local interactions, doesn't inherently optimize for or significantly disrupt global path lengths. It is interesting to note that Gemini, the LLM from Google, the company who shaped much of today's existing web structure, creates networks that are notably different from other LLMs in terms of compactness and clustering. ChatGPT and Claude, the other market leader in the LLM space, create networks with notably longer path lengths, while Llama (the LLM from Meta's Facebook and Instagram) creates networks with more moderate clustering.

These findings raise critical questions about the representational ethics and epistemic foundations of LLM-driven multi-agent AI systems. If language models replicate and even exaggerate social asymmetries without explicit instruction, then researchers and practitioners must ask not only what patterns are produced, but why and to what end. The directional biases in cross-category tie formation, for instance, may be symptomatic of broader cultural narratives encoded in training data, suggesting that these models inherit and reinstantiate social hierarchies embedded in language. Rather than treating LLMs as neutral engines of synthetic behavior, we might instead conceptualize them as socio-cognitive mirrors reflecting and amplifying the latent biases of the societies they were trained on. This carries implications not just for the study of artificial sociality, but also for our understanding of the subtle ways human prejudices may persist, mutate, or even be magnified in synthetic systems.

\section{Limitations and Outlook} 

This study has several limitations that warrant further exploration. First, to prioritize systematic tractability in this first exploration, our model specifications are relatively simplistic. The decision-making prompts employed in this framework fail to capture the full complexity of potential interactions, such as context-dependent decision-making, cultural norms, multiple attributes, and evolving biases. Additionally, our operationalization of attributes as binary characteristics oversimplifies the multifaceted nature of these traits. Furthermore, the LLM is treated as a static, memory-less decision-making entity, whereas, in reality, these models continuously evolve as they are exposed to new data. This dynamic adaptation could significantly influence long-term network evolution. Moreover, the restricted input size of LLMs necessitated the use of partial network information rather than a complete global view, impacting the network growth model.

Another significant limitation arises from our choice of LLMs. Given the computational demands (in order to generate our networks, we did more than $10^6$ API consultations, we used relatively simple LLMs. Those earlier and simple zero-shot models do not exhibit reasoning processes, and exhibit more hallucinations than other more advanced models. Specifically, we observed instances where the models generated node IDs that were deviated from the exact identifiers specified in the prompt. These hallucinations indicate a failure in the model’s ability to strictly adhere to task constraints, which is critical in controlled node selection settings~\cite{ji2023survey,manakul2023selfcheckgpt}. To address this issue, we initially attempted to mitigate hallucinations by re-querying the model multiple times to assess the consistency of its responses. However, when hallucinations persisted, we implemented a fallback procedure: we selected the closest matching node ID from the original list. While this workaround ensured the continuity of the experimental pipeline, it introduces a minor intervention that may compromise the interpretability and purity of the AI's autonomous decision-making process. Interestingly, it remains unclear whether humans exhibit analogous behavior—potentially making random choices during cognitive lapses—akin to these model hallucinations. Future work could address this limitation by exploring this phenomenon in greater depth.

A further limitation, and an important consideration for future work, is the potential influence of naming bias on our results. De Marzo et al.~\cite{de2023emergence} demonstrated that LLMs can exhibit a \textit{token prior}, showing a preference for certain randomly generated agent IDs over others, which can impact network formation if not controlled for. Our study did not incorporate such a control. However, in our primary attribute-based scenarios, the effect of this bias is likely secondary. Our results show that the introduction of social attributes triggers a powerful homophilic response, with assortativity coefficients for attributes like \textit{Political Orientation} exceeding $0.95$. This indicates that the decision-making process becomes dominated by the semantic meaning of the attributes rather than the syntax of the agent IDs. While the naming bias may have introduced a small amount of noise, the strength and statistical significance of our findings on homophily suggest that the social signal is overwhelmingly the primary driver in these cases. Future studies could integrate a renaming protocol to isolate the effect of social attributes with even greater precision, but the fact that our results emerged so strongly even in the potential presence of this confounding factor speaks to the robustness of homophily as a core principle in LLM-driven network generation.

While our result provide a robust view of the emergent structural patterns under our attachment mechanism, the generalizability to significantly larger networks (e.g., 10,000+ nodes) warrants further investigation. Although our model does not incorporate any notion of distance between nodes, scale effects can still influence global properties such as degree distribution or assortativity. For example, in larger networks, the preferential attachment process may lead to a more pronounced 'rich-get-richer' effect and potentially more stable power-law behavior. Similarly, assortativity patterns may shift due to the increasing diversity in node degrees. Future studies should explore whether these trends persist or change in larger-scale implementations.

Building upon these initial insights, future research should rigorously explore the dynamics of emergent biases over time and context, rather than simply their static manifestations. For instance, longitudinal simulations could investigate whether observed homophilic and asymmetric connection patterns intensify, dissipate, or reorganize as networks grow (e.g., with the introduction of new nodes) or are perturbed (e.g., through edge or node removals). Further, examining the impact of dynamic node attributes, where individual characteristics evolve over time, could reveal how LLM agents adapt their rewiring decisions to fluid identities. Beyond observation, a promising avenue involves active interventions to alter these emergent structures. This could include testing the efficacy of varied prompting strategies (e.g., explicit instructions for diversity, role-based prompting for inclusivity), exploring the potential of fine-tuning LLMs on specific datasets to promote desired connection patterns, or employing counterfactual inputs (e.g., hypothetically reversing attribute categories) to expose and understand latent biases in LLM decision-making. Finally, to ground these findings in practical relevance, future work should prioritize empirical validation and comparison to real-world analogues. Utilizing diverse empirical social datasets with comparable node attributes would allow for rigorous comparative analysis of network metrics (e.g., assortativity, community structure) and dynamic behaviors, ultimately discerning which aspects of human-like network evolution are faithfully reproduced by LLM agents and which are artifacts of model training or methodological design.

A significant and promising direction for future work is the development of an explicit mathematical model to formalize and reproduce the network formation behaviors we observed. The fact that some global properties like network compactness appear independent of node attributes, while local properties like clustering vary significantly, implies that the LLMs' decision-making follows a consistent, modelable logic. The ultimate goal would be to define the probability function, \( \Pr(v_i \mid S) \), which our study conceptualized. Based on our empirical results, this function would need to be a stochastic mechanism that balances two primary components: a structural preference component, which, based on our baseline analysis, seems to favor connections to the highest-degree nodes within the accessible subgroup, and a homophily preference component, which increases the connection probability for nodes sharing the same attribute, with the strength of this preference varying by attribute.

Finally, as autonomous AI agents increasingly shape the architecture of online systems, this work underscores the urgency of integrating ethical foresight into technical design. The emergent structures generated by LLM-based multi-agent systems reflect not only algorithmic logic but also embedded social assumptions. Understanding and steering these dynamics is essential to ensure that the networks built by AI reinforce inclusive, cohesive, and fair digital ecosystems rather than reproducing or exacerbating societal divides.

\section*{Statements and Declarations}

\textbf{Supplementary Information:} Supplementary material relevant to this study is provided in the appendix.

\textbf{Conflict of Interest:} The authors declare no known competing financial interests or personal relationships that could have influenced the work reported in this paper.

\textbf{Data Availability:} Datasets are available by contacting the authors.

\textbf{Code Availability:} Code used for network generation and analysis is available from the authors upon request.

\textbf{Author Contributions:} All authors contributed equally to the development and completion of this manuscript.

\bibliographystyle{unsrt}  
\bibliography{references}  

\appendix
\counterwithin{table}{section} 
\counterwithin{figure}{section}
\renewcommand\thesection{S.I.\arabic{section}}
\newpage

\section{Identifying Node Attributes}
\label{sec:S.I.1. Identifying Node Attributes}
\addcontentsline{toc}{section}{S.I.1. Identifying Node Attributes}

There is an indefinite number of possible node attributes we could have tested in our study. In order to identify the most prominent ones, we could have done a meta-review of the literature and identified a collection of commonly tested attributes. Going with the theme of an AI-agentic social fabric, we decided to consult AI instead. We consulted 10 different foundational Large Language Models (LLMs) from six different companies (Anthropic, Google, OpenAI, Meta, Mistral, X) about the most important network attributes to be considered in homophilic network dynamics. Since these models have been trained on unprecedented amounts of data (going beyond academic literature and including context from everyday dynamics collected by common crawl), and since transformer-based architectures are specialized to predict probabilities, we consider this to be a solid method to identifying the most probable candidates for relevant network attributes in our study. 
It is important to note that for this exercise, we used a context window that is independent from any exercise we did in the main study.

We gave each of the ten LLMs the following prompt:

\begin{quote}
When someone joins a social network, it is well known that homophily plays an important role in determining who to connect with ("birds of a feather flock together"). What are the most prominent node attributes usually considered by humans when looking for similar others? Make a list of the 10 most important ones and give a justification for each one. Then, rank them in priority to be typically considered by humans, from 1 (most important) to 10 (least important).
\end{quote}

We consulted the following LLMs: Llama 3.1-405B (Meta.AI), Mistral Large (Mistral AI), Gemini 1.5 Pro (Google), Gemini 1.5 Pro with Deep Research (Google), Gemini 2.0 Pro (Google), ChatGPT4 (OpenAI), ChatGPT-01 (OpenAI), Claude 3.5 Sonnet (Anthropic), Claude 3 Opus (Anthropic). 

Below we present the full list of their responses, including URLs.

It was straightforward to group every one of the $10 \times 10 = 100$ mentions of prominent node attributes into 17 distinct groups, with, on average, 5.9 mentions per group (see Table S.I.1). The node attributes \textbf{[Interest/Hobbies]}, \textbf{[Age]}, \textbf{[Education]}, and \textbf{[Professional]} were mentioned by all 10 LLMs. We calculated the average rank of attributes for each group based on the data in the fourth column of Table S.I.1. Additionally, we computed another average assuming there were 10 topics per group. For this calculation, we assigned a rank of 11 to any topic that was not mentioned, under the assumption that each group should contain 10 topics (column 5 in Table S.I.1).

\begin{table}[h]
    \centering
    \caption{Summary Table of mentioned homophilic node attributes by 10 LLMs}
    \resizebox{\textwidth}{!}{
        \begin{tabular}{clcccc}
            \toprule
            \textbf{Rank} & \textbf{Attribute} & \textbf{Number of Mentions} & \textbf{Simple Avg. of Ranks} & \textbf{Expanded Avg. Non-Mentions = Rank 11} & \textbf{Overall Avg. Rank} \\
            \midrule
            1 & Interests / Hobbies & 10 & 2.2 & 3.7 & 2.85 \\
            2 & Age & 10 & 2.3 & 3.8 & 2.94 \\
            3 & Education & 10 & 4.5 & 5.6 & 4.96 \\
            4 & Location & 9 & 4.4 & 6.1 & 5.14 \\
            5 & Values \& Beliefs & 4 & 3.0 & 8.3 & 5.33 \\
            6 & Professional & 10 & 5.6 & 6.5 & 5.97 \\
            7 & Cultural & 8 & 6.4 & 7.9 & 6.98 \\
            8 & Ethnicity & 4 & 6.0 & 9.3 & 7.33 \\
            9 & Political & 6 & 7.3 & 9.2 & 8.00 \\
            10 & Religion & 6 & 7.5 & 9.3 & 8.13 \\
            11 & Gender & 4 & 7.3 & 9.8 & 8.17 \\
            12 & Personality & 4 & 7.3 & 9.8 & 8.17 \\
            13 & Family Status & 4 & 8.0 & 10.0 & 8.67 \\
            14 & Socio-economic Status & 4 & 8.0 & 10.0 & 8.67 \\
            15 & Lifestyle & 3 & 8.7 & 10.4 & 9.17 \\
            16 & Appearance & 2 & 10.0 & 10.8 & 10.00 \\
            17 & Leisure & 2 & 10.0 & 10.8 & 10.00 \\
            \bottomrule
        \end{tabular}
    }
    \label{tab:homophilic_attributes}
\end{table}

\begin{figure}[ht]
    \centering
    \includegraphics[width=0.8\textwidth]{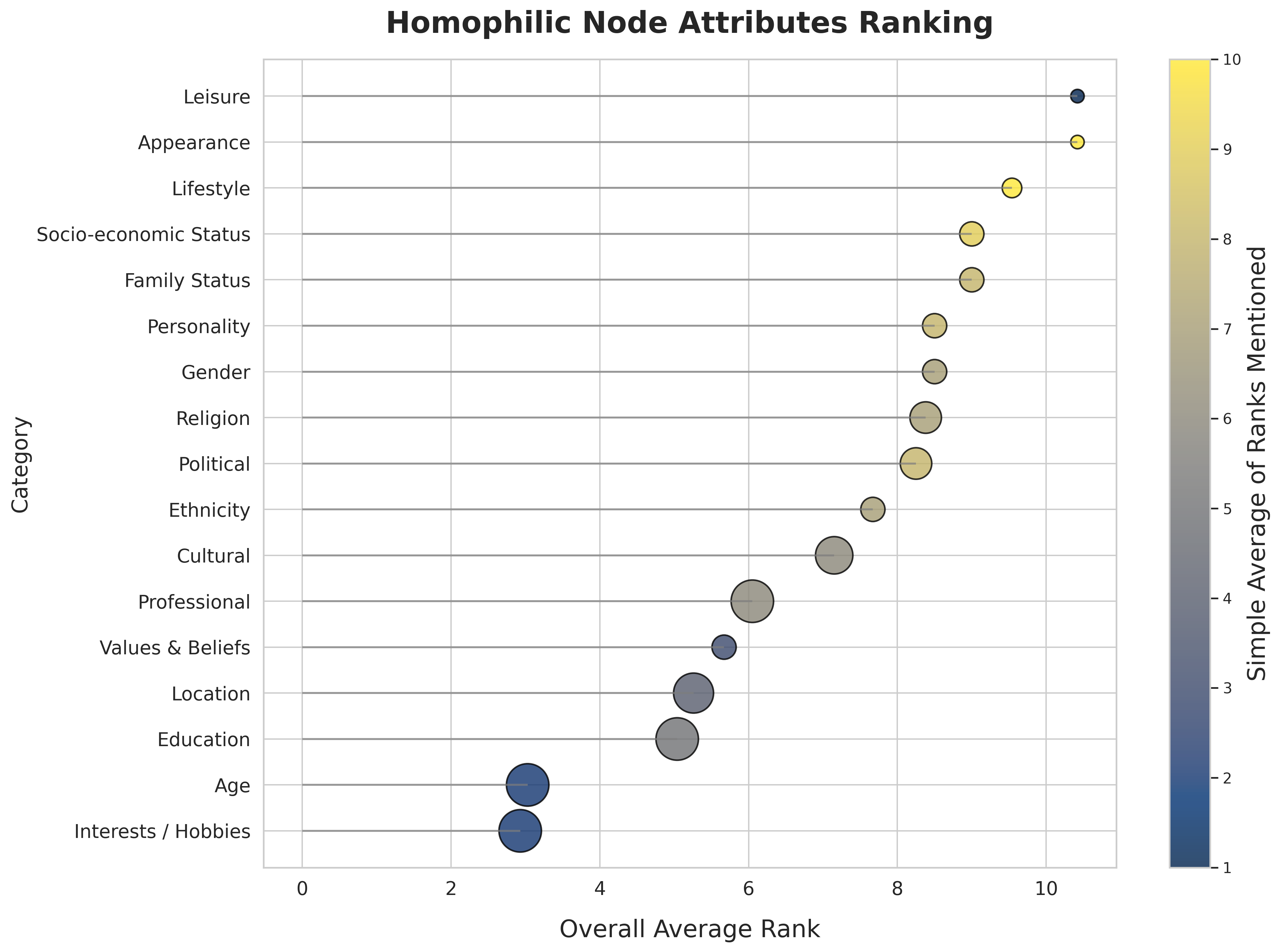}
    \caption{Horizontal lollipop chart showing the average ranks of the most prominent homophilic node attributes. Each category is represented by a bubble, with the horizontal position indicating the overall average rank and the color representing the simple average of ranks mentioned. Bubble size encodes the number of mentions (min = 2, max = 10).}
    \label{fig:scatter_plot}
\end{figure}

Given computational costs, we decided to go with half of the 17 possible topics, resulting in 8 groups. Considering their potential prominence, we decided to include [Interest/Hobbies], [Age], and [Education] into our study. Given their social relevance and legal protection, we decided to include [Gender], [Ethnicity], [Religion], [Political], and [Socio-economic status] into our study. This leaves us with 8 attributes to be tested.

Keeping combinational complexity to a minimum, we created binary variables for each of them. We did so by feeding the provided definitions and explanations of the LLMs (see below) into Claude 3.5 Sonnet (Anthropic) and asked it to create a binary distinction for each.

Building on our previous prompt, we gave Claude the following prompt (with XXX being the group), which led to the binary attributes presented in Table S.I.2:

\begin{quote}
    When someone joins a social network, it is well known that homophily plays an important role in determining who to connect with ("birds of a feather flock together"). Below I give you some justifications why XXX matters in that choice. Based on this, please come up with the most prominent binary distinction of XXX that can influence the choice of someone joining the network. Here are the justifications:
\end{quote}

\begin{table}[h]
\centering
\renewcommand{\arraystretch}{1.3}
\begin{tabular}{|l|c|c|}
\hline
\textbf{Attribute} & \textbf{Category 1} & \textbf{Category 2} \\
\hline
Gender & Female & Male \\
Age & Under 30 years old & Over 30 years old \\
Political Orientation & Liberal & Conservative \\
Religion & Religious (practicing) & Secular (non-practicing) \\
Interests/Hobbies & Active Engagement & Passive Engagement \\
Education & College Degree & No College Degree \\
Ethnicity & Western & Non-Western \\
Socio-economic Status & Professional/Managerial Class & Working Class \\
\hline
\end{tabular}
\caption{Binary attribute classifications.}
\label{table:binary_attributes_app}
\end{table}

\subsection{Grok (by X, formerly Twitter)}
\addcontentsline{toc}{subsection}{S.I.1.1 Grok (by X, formerly Twitter)}
\url{https://x.com/i/grok/share/ALR4bO8n6KBPlIMa46lPJD6qj}

\textbf{Ranking by Priority:}
\begin{enumerate}
    \item Interests/Hobbies - Most fundamental for initiating and maintaining conversation and engagement.
    \item Age - Critical for understanding life stage and potential for shared experiences.
    \item Geographical Location - Essential for practical reasons in both online and offline contexts.
    \item Profession/Occupation - Provides a basis for meaningful professional and personal connections.
    \item Educational Background - Influences perspectives and can be a significant bonding factor.
    \item Cultural Background - Affects daily life, communication styles, and social norms.
    \item Political Views - Can dictate the comfort level in discussions and community involvement.
    \item Religion/Spirituality - Core for some individuals in forming deep connections.
    \item Lifestyle Choices - Becoming increasingly important with the rise of specific lifestyle communities.
    \item Appearance/Aesthetics - Least influential in forming deep connections but can affect initial attraction or group dynamics.
\end{enumerate}

\subsection{Llama 3.1-405B (Meta.AI)}
\addcontentsline{toc}{subsection}{S.I.1.2 Llama 3.1-405B (Meta.AI)}
\url{https://meta.ai/s/X2N1TrAHgWBabBqF/}

\textbf{Top 10 Node Attributes}
\begin{enumerate}
    \item Shared Interests/Hobbies (Priority: 1) - Humans tend to connect with others who share similar passions, hobbies, or interests. This attribute helps create a sense of community and fosters meaningful interactions.
    \item Demographics (Age, Location, Occupation) (Priority: 2) - People often seek connections with others from similar demographic backgrounds, as they can relate to shared life experiences, cultural norms, and socioeconomic contexts.
    \item Personality Traits (e.g., introvert/extrovert, optimistic/pessimistic) (Priority: 3) - Individuals may seek connections with others who share similar personality characteristics, facilitating easier communication and understanding.
    \item Values and Beliefs (e.g., political, religious, moral) (Priority: 4) - Shared values and beliefs can create strong bonds between individuals, as they often influence worldview, behavior, and decision-making.
    \item Language and Culture (Priority: 5) - Language and cultural background can significantly impact communication and relationships. People often seek connections with others who share their native language and cultural heritage.
    \item Education and Socioeconomic Status (Priority: 6) - Similar educational backgrounds and socioeconomic statuses can create common ground and facilitate understanding of shared experiences and challenges.
    \item Lifestyle (e.g., health-conscious, family-oriented, career-driven) (Priority: 7) - Individuals may seek connections with others who share similar lifestyle choices, habits, or priorities, promoting mutual support and understanding.
    \item Hometown or Geographic Origin (Priority: 8) - People often connect with others from their hometown or geographic region, sharing nostalgia, local knowledge, and cultural heritage.
    \item Professional Affiliations or Industry (Priority: 9) - Shared professional backgrounds or industry affiliations can create opportunities for networking, collaboration, and knowledge sharing.
    \item Leisure Activities (e.g., sports, music, travel) (Priority: 10) - While important, leisure activities are often considered secondary to other attributes, as they may not be as central to one's identity or daily life.
\end{enumerate}

\subsection{Mistral Large (Mistral AI)}
\addcontentsline{toc}{subsection}{S.I.1.3 Mistral Large (Mistral AI)}
\url{https://chat.mistral.ai/chat/b1956eb7-cee3-4241-b3c1-2759dc5d8f16}

\textbf{Ranking in Priority:}
\begin{enumerate}
    \item Interests/Hobbies: Shared interests are often the most immediate and visible attributes that drive initial connections.
    \item Location: Proximity is crucial for face-to-face interactions and shared local experiences.
    \item Age: Similar age groups often have comparable life stages and experiences.
    \item Occupation/Industry: Shared professional backgrounds can lead to meaningful connections and collaborations.
    \item Education Level: Similar educational backgrounds can foster intellectual and career-related connections.
    \item Gender: Shared gender experiences can influence social dynamics and connections.
    \item Cultural Background: Shared cultural values and traditions can create a sense of community and belonging.
    \item Relationship Status: Similar relationship statuses can lead to shared life experiences and support.
    \item Language: Speaking the same language facilitates communication and understanding.
    \item Political Views: While important, political views may not be as immediately visible or influential in initial connections compared to other attributes.
\end{enumerate}

\subsection{Gemini 1.5 Pro (Google)}
\addcontentsline{toc}{subsection}{S.I.1.4 Gemini 1.5 Pro (Google)}
\url{https://g.co/gemini/share/f498f275a5cc}

\textbf{Here's a list of 10 prominent node attributes people consider, ranked in typical order of importance:}
\begin{enumerate}
    \item Shared Interests: This is often the most crucial factor. People connect with those who enjoy similar activities, hobbies, or have passions in common (e.g., sports, music, books, movies, etc.). This provides a basis for conversation, shared experiences, and potential activities together.
    \item Values and Worldview: Shared values and beliefs about important issues form a strong foundation for connection. This can include political or religious affiliations, social causes they support, or general outlook on life.
    \item Age or Life Stage: People tend to gravitate towards others in similar age brackets or life stages (e.g., students, young professionals, parents, retirees). This leads to shared experiences, common challenges, and relatable perspectives.
    \item Background and Culture: Shared cultural background, ethnicity, or upbringing can create an immediate sense of connection. This can involve language, traditions, values, and understanding of specific cultural nuances.
    \item Education Level: People often connect with others who have similar educational attainment. This can lead to common interests, intellectual conversations, and understanding of professional challenges.
    \item Occupation or Career Field: Those in the same or related professions often form connections due to shared experiences, industry knowledge, and professional networks.
    \item Geographic Proximity: While less crucial in online networks, location still plays a role. People often connect with those living in the same city, region, or country, facilitating in-person interactions and shared local experiences.
    \item Social Status or Class: Though often implicit, people tend to connect with others of similar socioeconomic backgrounds. This can involve shared lifestyles, access to resources, and understanding of social norms.
    \item Personality Traits: Shared personality traits like introversion/extroversion, sense of humor, or openness to new experiences can contribute to compatibility and enjoyment of each other's company.
    \item Physical Appearance: While perhaps superficial, physical attractiveness and perceived similarity in appearance can play a role in initial attraction and connection, particularly in online settings.
\end{enumerate}

\subsection{Gemini 1.5 Pro with Deep Research (Google)}
\url{https://g.co/gemini/share/503b1a38e0b7}

\textbf{Ranked Attributes and Justifications:}
\begin{enumerate}
    \item \textbf{Values and Beliefs} - Shared values and beliefs foster a sense of understanding, trust, and common ground, which are essential for building strong relationships. These can include moral principles, religious convictions, political ideologies, and worldviews. For example, individuals with similar political affiliations or attitudes towards social issues like climate change or gender equality might be more likely to connect. Value homophily strengthens relationships and facilitates collective identity and action.
    \item \textbf{Ethnicity} - Ethnicity plays a significant role in shaping social identity and influencing social connections. Individuals often gravitate towards those with shared ethnic backgrounds due to cultural similarities, shared experiences, and a sense of belonging. This can be seen in the formation of ethnic communities and social groups within larger networks. Research indicates that ethnicity can be a strong driver of inbreeding homophily, especially in diverse societies. It's important to note that ethnicity often intersects with other attributes like socioeconomic status and cultural background, further influencing connection patterns.
    \item \textbf{Age} - Age similarity promotes shared experiences, common interests, and easier communication. People of similar ages often go through similar life stages and developmental milestones, leading to greater understanding and empathy. This can be seen in the formation of age-based peer groups and social circles. Age homophily is primarily baseline, with individuals connecting with those in their age group due to shared environments and social circles. However, age can also interact with other attributes, such as hobbies and interests, with individuals of similar ages being more likely to engage in similar activities.
    \item \textbf{Gender} - Gender plays a fundamental role in shaping social interactions and friendships. Individuals often feel more comfortable and understood among those who share their gender identity. This can be attributed to shared experiences, common interests, and social norms related to gender roles. Gender homophily is influenced by societal norms and expectations, with men and women sometimes exhibiting different friendship patterns and communication styles. For example, women's friendships might emphasize emotional intimacy and self-disclosure, while men's friendships might focus on shared activities and competition.
    \item \textbf{Religion} - Religious beliefs often influence values, morals, and lifestyles. Individuals with shared religious affiliations may find common ground in their spiritual practices, beliefs, and community involvement. This can be seen in the formation of religious communities and social groups within networks. Religious homophily can be attributed to both baseline and inbreeding homophily, with individuals seeking connections within their religious groups for social support and shared values. Religious beliefs can also interact with other attributes like values and ethnicity, further influencing connection patterns.
    \item \textbf{Education} - Educational background can shape perspectives, knowledge, and social circles. Individuals with similar educational attainment may share common interests, intellectual pursuits, and career paths. This can be seen in the formation of alumni networks and professional communities. Educational homophily can be influenced by institutional structures and social contexts, such as attending the same school or university. Educational background can also interact with other attributes like profession and income, influencing social mobility and network connections.
    \item \textbf{Profession} - Shared professional experiences and career paths can foster understanding, common interests, and networking opportunities. Individuals in similar professions may face similar challenges, share industry knowledge, and provide mutual support. This can be seen in the formation of professional associations and industry-specific networks. Professional homophily can be driven by workplace interactions, shared goals, and professional identity. Professional background can also interact with other attributes like education and income, influencing career progression and social status.
    \item \textbf{Hobbies and Interests} - Shared hobbies and interests provide a foundation for common activities, conversations, and social connections. Individuals who engage in similar leisure activities often find enjoyment and camaraderie in shared passions. This can be seen in the formation of hobby groups, fan communities, and online forums. Hobbies and interests can contribute to value homophily, as individuals with similar interests often share underlying values and preferences. For example, individuals who enjoy outdoor activities might also share values related to environmentalism and sustainability.
    \item \textbf{Income} - Income level can influence lifestyle, social circles, and access to resources. Individuals with similar income levels may share similar socioeconomic backgrounds, consumption patterns, and social experiences. This can be seen in the formation of social classes and economic strata within networks. Income homophily can be influenced by residential segregation, social stratification, and access to social and economic opportunities. Income can also interact with other attributes like education and profession, influencing social mobility and network connections.
    \item \textbf{Lifestyle} - Lifestyle encompasses various aspects of daily life, including habits, preferences, and social activities. Individuals with similar lifestyles may find common ground in their daily routines, leisure pursuits, and social interactions. This can be seen in the formation of communities based on shared lifestyles, such as fitness enthusiasts, foodies, or travelers. Lifestyle homophily can be influenced by shared values, interests, and social circles. For example, individuals with similar health-conscious lifestyles might also share values related to well-being and personal growth.
\end{enumerate}

\subsection{Gemini 1.5 Pro with Deep Research (Google)}
\addcontentsline{toc}{subsection}{S.I.1.5 Gemini 1.5 Pro with Deep Research (Google)}

\textbf{1. Shared Interests and Hobbies} \\
\textit{Justification:} This is the most immediate and visible point of connection. Shared interests provide common ground for conversation, activities, and potential friendship. Whether it is playing the same sport, having an affinity for the same musical genre, or a shared passion for hiking, this is a significant factor.

\textbf{2. Age/Generation} \\
\textit{Justification:} People in the same age cohort often share similar life experiences, cultural references, and developmental stages. This can lead to a sense of understanding and shared identity.

\textbf{3. Geographic Proximity} \\
\textit{Justification:} While online networks transcend physical boundaries, proximity still matters. People who live near each other are more likely to encounter one another in offline settings, share local experiences, and have opportunities for face-to-face interaction.

\textbf{4. Educational Background} \\
\textit{Justification:} Similar education levels often imply shared intellectual interests, values, and socioeconomic status. Attending the same school or studying the same subject can be a strong bonding factor.

\textbf{5. Values and Beliefs} \\
\textit{Justification:} Shared values (e.g., political views, religious beliefs, environmental concerns) are fundamental to our sense of self and how we interact with the world. Connecting with others who share these core values creates a sense of validation and community.

\textbf{6. Social Status and Lifestyle} \\
\textit{Justification:} People tend to gravitate towards others with similar social standing and lifestyles. This can be driven by shared aspirations, comfort levels, and a desire for social validation.

\textbf{7. Race and Ethnicity} \\
\textit{Justification:} While it's unfortunate that this remains a factor, shared racial or ethnic background can lead to a sense of cultural affinity, shared history, and potential understanding of lived experiences, particularly in the face of discrimination.

\textbf{8. Professional Network/Industry} \\
\textit{Justification:} Working in the same field or industry suggests shared career goals, experiences, and knowledge. This can be a powerful driver of connection, particularly for professional networking and mentorship.

\textbf{9. Family Background/Structure} \\
\textit{Justification:} Coming from similar family backgrounds (e.g., marital status of parents, number of siblings, family values) can create a sense of relatability and shared understanding of family dynamics. Similar family structures (e.g., single, married with children) can also lead to common interests related to family life.

\textbf{10. Personality Traits} \\
\textit{Justification:} While harder to assess initially in a social network, shared personality traits (e.g., introversion/extroversion, openness to experience) can influence compatibility and long-term friendship potential.

\subsection{ChatGPT4 (OpenAI)}
\addcontentsline{toc}{subsection}{S.I.1.7 ChatGPT4 (OpenAI)}

\textbf{1. Interests and Hobbies} \\
\textit{Justification:} Shared interests and hobbies form the basis of many social interactions and conversations, making them key factors in establishing and maintaining connections.

\textbf{2. Location (Geographic Proximity)} \\
\textit{Justification:} People often connect with others who are geographically close to facilitate face-to-face interactions and participate in local events or communities.

\textbf{3. Age} \\
\textit{Justification:} Age affects life stages, interests, and social circles, influencing how individuals relate to one another and the topics they discuss.

\textbf{4. Educational Background} \\
\textit{Justification:} Similar educational levels or institutions can lead to shared experiences, networks, and a common understanding of certain topics, fostering deeper connections.

\textbf{5. Political Views} \\
\textit{Justification:} Political affiliations can significantly influence a person's worldview and discussions, often leading individuals to seek out others who share similar political perspectives.

\textbf{6. Ethnicity/Cultural Background} \\
\textit{Justification:} Shared cultural or ethnic backgrounds can enhance connectivity through common languages, traditions, and experiences.

\textbf{7. Professional Field/Career} \\
\textit{Justification:} People in similar professions or industries tend to connect for networking, professional development, and discussing industry-specific topics.

\textbf{8. Religion} \\
\textit{Justification:} Religion can be a core aspect of identity and values, leading individuals to seek connections with those who have similar beliefs and practices.

\textbf{9. Gender} \\
\textit{Justification:} Gender can influence socialization patterns and interests, potentially making individuals more likely to connect with those of the same gender.

\textbf{10. Sexual Orientation} \\
\textit{Justification:} Shared experiences related to sexual orientation can form the basis for community support, understanding, and shared challenges or triumphs.

\subsection{ChatGPTo1 (OpenAI)}
\addcontentsline{toc}{subsection}{S.I.1.8 ChatGPTo1 (OpenAI)}

\textbf{1. Shared Interests and Hobbies} \\
\textit{Justification:} People are naturally drawn to those who share their passions—be it a love of sports, music, gaming, art, or any other pastime. Common hobbies provide an immediate conversation starter and a sense of community or belonging. \\
\textit{Why High Priority?} In many online communities (e.g., Facebook groups, Reddit subs), users often connect primarily because of a shared interest or topic.

\textbf{2. Age or Life Stage} \\
\textit{Justification:} Individuals in the same age range (or life stage) often share cultural references, life milestones (e.g., entering college, starting a family), and generational values. This creates a natural sense of understanding and rapport. \\
\textit{Why High Priority?} People in similar age brackets tend to relate to one another’s experiences and challenges (e.g., recent graduates bonding over job hunts, retirees sharing travel plans).

\textbf{3. Education Level or Institutional Affiliation} \\
\textit{Justification:} Humans often form social circles around schools, universities, and alumni groups. Shared educational backgrounds can hint at similar cognitive approaches, values, and networks. \\
\textit{Why High Priority?} Whether bonding over a shared major or the prestige of an alma mater, educational ties tend to foster strong social and professional connections.

\textbf{4. Occupation or Professional Field} \\
\textit{Justification:} Occupations shape daily routines, skill sets, and viewpoints. People who work in the same industry or profession often face similar challenges and opportunities, making it natural to connect and share knowledge. \\
\textit{Why High Priority?} Professional networking sites (e.g., LinkedIn) exist largely because career-based homophily is a strong basis for connection and mutual benefit.

\textbf{5. Geographic Proximity (Location)} \\
\textit{Justification:} People who live or have lived in the same area tend to share local culture, dialects, and experiences. Proximity is a classic driver of social ties—both offline and online (via location-based groups). \\
\textit{Why High Priority?} Local ties can mean shared schools, social events, or local community issues, leading to a sense of familiarity and trust.

\textbf{6. Language} \\
\textit{Justification:} Being able to communicate smoothly is fundamental. Shared language not only removes communication barriers but also implies shared cultural references (slang, humor, idioms). \\
\textit{Why High Priority?} Even in global social networks, language-based communities are common; users tend to cluster by shared language to interact more comfortably.

\textbf{7. Personality or Temperament} \\
\textit{Justification:} Beyond superficial traits, people often seek others with comparable energy levels, emotional styles, or even sense of humor (e.g., introverts gravitating to similarly reserved individuals). \\
\textit{Why Medium Priority?} Personality similarity is critical for long-term close friendships, though it can be less obvious online initially compared to more explicit attributes like interests or age.

\textbf{8. Religion or Spiritual Beliefs} \\
\textit{Justification:} Shared faith can be a powerful connector, offering a deep sense of identity and community. Religious groups often serve as social hubs, both locally and online. \\
\textit{Why Medium Priority?} While religion can profoundly shape values and lifestyle, not everyone prioritizes it when seeking connections. However, in faith-based networks or communities, it becomes highly salient.

\textbf{9. Ethnicity or Cultural Background} \\
\textit{Justification:} Shared heritage or cultural background often implies similar traditions, customs, and social norms. This can lead to easier rapport and understanding of cultural references. \\
\textit{Why Lower Priority?} Ethnic or cultural ties can be very strong in some contexts but are less explicitly pursued in diverse, global online communities—unless the purpose is to celebrate a particular culture or identity.

\textbf{10. Political Orientation or Ideological Views} \\
\textit{Justification:} People frequently gravitate toward those who share (or at least don’t oppose) their political or ideological stances. This is especially evident in politically charged social media groups. \\
\textit{Why Lower Priority?} While politically oriented communities thrive, many casual social seekers often sidestep overt political discussions initially, focusing on less divisive attributes like hobbies or location.

\subsection{Claude 3.5 Sonnet (Anthropic)}
\addcontentsline{toc}{subsection}{S.I.1.9 Claude 3.5 Sonnet (Anthropic)}

\textbf{Ranked by typical priority in human decision-making for social connections:}

\textbf{1. Age (Most fundamental and immediately visible)} \\
\textit{Justification:} Age is one of the most immediately visible factors in establishing connections. It often reflects shared generational experiences and cultural references, which facilitate social bonding.

\textbf{2. Geographic Location (Practical necessity for real-world interaction)} \\
\textit{Justification:} Proximity is important for in-person interactions, shared local experiences, and accessibility to events and activities. Geographic location influences the possibility of offline connections.

\textbf{3. Professional Field (Critical for career advancement and daily life)} \\
\textit{Justification:} Common professions or industries help establish rapport and mutual understanding regarding professional experiences, work-related challenges, and career growth opportunities.

\textbf{4. Educational Background (Strong indicator of shared experiences and knowledge)} \\
\textit{Justification:} Similar educational paths can create strong bonds, as individuals from similar institutions or fields often share values, cognitive frameworks, and knowledge bases.

\textbf{5. Cultural Background (Important for communication and shared values)} \\
\textit{Justification:} Cultural background shapes individuals’ worldviews, norms, and ways of communication, playing a key role in understanding and compatibility in relationships.

\textbf{6. Interests/Hobbies (Creates opportunities for enjoyable interaction)} \\
\textit{Justification:} Shared hobbies or passions provide natural opportunities for enjoyable interactions and mutual enjoyment, helping to strengthen connections and create lasting bonds.

\textbf{7. Political Views (Increasingly important but not always primary)} \\
\textit{Justification:} Political alignment often influences social interactions, as individuals tend to gravitate toward others who share their ideological stances, especially in polarized environments.

\textbf{8. Family Status (Relevant but not always decisive)} \\
\textit{Justification:} Family status can influence lifestyle, priorities, and expectations in relationships. However, it may not always play a decisive role in forming connections, especially outside family-oriented communities.

\textbf{9. Socioeconomic Status (Influential but often implicit)} \\
\textit{Justification:} Socioeconomic status can subtly influence compatibility, as it can reflect access to resources, education, and opportunities. While important, it is often a less overt factor in social decisions.

\textbf{10. Religious/Spiritual Beliefs (Important for some, less so for others)} \\
\textit{Justification:} Religion or spiritual beliefs can form a foundational aspect of identity for some people, creating deep connections with others who share the same faith or values.

\subsection{Claude 3 Opus (Anthropic)}
\addcontentsline{toc}{subsection}{S.I.1.10 Claude 3 Opus (Anthropic)}

\textbf{1. Shared interests and hobbies} \\
\textit{Justification:} People naturally seek out others who enjoy the same activities, as it provides common ground for interaction and relationship building.

\textbf{2. Age} \\
\textit{Justification:} Individuals tend to connect more with people in a similar stage of life, as they can relate to shared experiences and challenges.

\textbf{3. Education level} \\
\textit{Justification:} Those with comparable educational backgrounds often share similar values, knowledge, and socioeconomic status, fostering connections.

\textbf{4. Occupation and industry} \\
\textit{Justification:} Connecting with others in the same professional field allows for networking, career growth, and mutual understanding of work-related topics.

\textbf{5. Political beliefs} \\
\textit{Justification:} In an increasingly polarized world, people often seek out others who align with their political views to avoid conflict and find support.

\textbf{6. Religious beliefs} \\
\textit{Justification:} Shared faith provides a strong foundation for building relationships and engaging in religious communities online.

\textbf{7. Marital status and family structure} \\
\textit{Justification:} People in similar relationship and family situations (e.g., single, married, parents) can relate to each other's lifestyles and priorities.

\textbf{8. Geographic location} \\
\textit{Justification:} Connecting with others nearby allows for the possibility of offline interactions and fosters a sense of local community.

\textbf{9. Language and cultural background} \\
\textit{Justification:} Shared language and cultural heritage provides a sense of familiarity and belonging, facilitating communication and understanding.

\textbf{10. Music, movie, and entertainment preferences} \\
\textit{Justification:} While less critical than other factors, bonding over favorite bands, genres, or TV shows is a common way to establish initial connections.

\newpage
\section{Degree and Attribute Assortativity Across Models and Attributes}
\addcontentsline{toc}{section}{S.I.2. Degree and Attribute Assortativity Across Models and Attributes}
\label{sec:assortativity_table}

\begin{table}[htbp]
\centering
\caption{Mean and Standard Error of Degree and Attribute Assortativity Across Models and Attributes}
\begin{tabular}{llrrrr}
\toprule
\textbf{Model} & \textbf{Attribute} & \textbf{Deg. Assort. (Mean)} & \textbf{SED} & \textbf{Attr. Assort. (Mean)} & \textbf{SED} \\
\midrule
\multirow{9}{*}{Gemini 1.5 Flash} 
& Degree-exclusive       & -0.516 & 0.0081 & -0.0035 & 0.0121 \\
& Political Orientation  & -0.422 & 0.0097 & 0.9632 & 0.0023 \\
& Gender                 & -0.475 & 0.0070 & 0.2444 & 0.0384 \\
& Age                    & -0.487 & 0.0091 & 0.7065 & 0.0068 \\
& Education              & -0.512 & 0.0044 & 0.3501 & 0.0253 \\
& Ethnicity              & -0.531 & 0.0084 & 0.3720 & 0.0198 \\
& Religious Practice     & -0.473 & 0.0073 & 0.6462 & 0.0223 \\
& Socio-Economic         & -0.521 & 0.0088 & 0.5775 & 0.0147 \\
& Interest               & -0.477 & 0.0095 & 0.2772 & 0.0615 \\
\midrule
\multirow{9}{*}{GPT-4o Mini} 
& Degree-exclusive       & -0.499 & 0.0091 & -0.0047 & 0.0123 \\
& Political Orientation  & -0.244 & 0.0084 & 0.9785 & 0.0015 \\
& Gender                 & -0.396 & 0.0089 & 0.8462 & 0.0104 \\
& Age                    & -0.297 & 0.0053 & 0.9699 & 0.0021 \\
& Education              & -0.184 & 0.0104 & 0.9078 & 0.0138 \\
& Ethnicity              & -0.276 & 0.0028 & 0.9747 & 0.0014 \\
& Religious Practice     & -0.178 & 0.0075 & 0.9113 & 0.0075 \\
& Socio-Economic         & -0.228 & 0.0079 & 0.9634 & 0.0013 \\
& Interest               & -0.323 & 0.0140 & -0.2241 & 0.0317 \\
\midrule
\multirow{9}{*}{Claude 3.0 Haiku}
& Degree-exclusive       & -0.504 & 0.0029 & 0.0029 & 0.0073 \\
& Political Orientation  & -0.211 & 0.0077 & 0.9702 & 0.0041 \\
& Gender                 & -0.398 & 0.0096 & 0.5105 & 0.0262 \\
& Age                    & -0.337 & 0.0085 & 0.9693 & 0.0034 \\
& Education              & -0.375 & 0.0098 & 0.4926 & 0.0192 \\
& Ethnicity              & -0.340 & 0.0064 & 0.8956 & 0.0099 \\
& Religious Practice     & -0.267 & 0.0086 & 0.7092 & 0.0233 \\
& Socio-Economic         & -0.286 & 0.0119 & 0.5412 & 0.0231 \\
& Interest               & -0.391 & 0.0160 & -0.3208 & 0.0140 \\
\midrule
\multirow{9}{*}{Llama-4-Scout}
& Degree-exclusive       & -0.465 & 0.0069 & -0.0246 & 0.0075 \\
& Political Orientation  & -0.358 & 0.0085 & 0.9397 & 0.0052 \\
& Gender                 & -0.389 & 0.0067 & 0.0067 & 0.0142 \\
& Age                    & -0.360 & 0.0070 & 0.6546 & 0.0109 \\
& Education              & -0.326 & 0.0178 & 0.5835 & 0.0193 \\
& Ethnicity              & -0.380 & 0.0049 & 0.7058 & 0.0110 \\
& Religious Practice     & -0.336 & 0.0107 & 0.7247 & 0.0243 \\
& Socio-Economic         & -0.342 & 0.0051 & 0.7992 & 0.0124 \\
& Interest               & -0.357 & 0.0057 & 0.0079 & 0.0214 \\
\bottomrule
\end{tabular}
\label{tab:assortativity_summary}
\end{table}

\subsection{Ambiguity in the \textit{Interest} Attribute and Exclusion from Aggregate Analysis}

In our original design, we included an attribute labeled \textit{interest}, distinguishing between categories that require ``active engagement'' and those involving ``passive engagement'' (see Table~\ref{tab:assortativity_summary}). However, we found that this classification introduced considerable ambiguity, leading to highly inconsistent mixing patterns across models. This suggests that LLMs may interpret the concept of \textit{interest} in different ways, raising concerns about construct validity.

To explore this further, we considered various interpretive lenses through which ``active'' and ``passive'' interests might be understood. These include:

\begin{itemize}
    \item \textbf{Physical vs. Sedentary}: Activities requiring physical movement (e.g., sports) vs. stationary ones (e.g., watching movies).
    \item \textbf{Cognitive Demand}: Tasks that involve mental effort (e.g., coding) vs. relaxation (e.g., casual browsing).
    \item \textbf{Production vs. Consumption}: Creating content (e.g., writing, cooking) vs. consuming it (e.g., reading, watching).
    \item \textbf{Social Engagement}: Interests involving interaction (e.g., team sports) vs. solitary or observational ones.
    \item \textbf{Agency and Control}: Activities with self-directed structure (e.g., teaching, entrepreneurship) vs. those following preset paths.
    \item \textbf{Skill Development}: Interests focused on mastery and practice (e.g., music, martial arts) vs. low-pressure enjoyment.
\end{itemize}

Interestingly, many interests can shift between ``active'' and ``passive'' depending on how they are approached. For instance, photography can be a passive hobby or an active, skill-based pursuit. Given this interpretive flexibility and the variability in LLM responses, we determined that \textit{interest} lacks the clarity required for reliable comparison across models and excluded it from our results.

\newpage
\section{Network Structure Metrics}
\addcontentsline{toc}{section}{S.I.3. Table of Network Structure Metrics}

\begin{table}[h!]
\centering
\caption{Summary of network structure metrics (average clustering coefficient, average shortest path length, and average diameter) across models and attributes.}
\label{tab:network-metrics-summary}
\resizebox{\textwidth}{!}{
\begin{tabular}{llccc}
\toprule
\textbf{Model} & \textbf{Attribute} & \textbf{Avg. Clustering} & \textbf{Avg. Shortest Path Length} & \textbf{Avg. Diameter} \\
\midrule
\multirow{9}{*}{GPT-4o mini}
 & Age & 0.28 & 3.17 & 5.83 \\
 & Religious & 0.16 & 3.33 & 6.00 \\
 & Socio-Economic & 0.20 & 3.29 & 6.33 \\
 & Ethnicity & 0.26 & 3.23 & 6.17 \\
 & Gender & 0.40 & 2.67 & 4.83 \\
 & Political & 0.22 & 3.30 & 6.17 \\
 & Interest & 0.23 & 2.78 & 5.00 \\
 & Degree-Exclusive & 0.47 & 2.50 & 4.00 \\
 & Education & 0.16 & 3.32 & 6.17 \\
\midrule
\multirow{9}{*}{Gemini 1.5 Flash}
 & Age & 0.45 & 2.70 & 5.00 \\
 & Religious & 0.39 & 2.69 & 5.00 \\
 & Socio-Economic & 0.46 & 2.61 & 4.75 \\
 & Ethnicity & 0.47 & 2.55 & 4.50 \\
 & Gender & 0.45 & 2.55 & 4.50 \\
 & Political & 0.41 & 3.00 & 5.33 \\
 & Interest & 0.35 & 2.72 & 5.00 \\
 & Degree-Exclusive & 0.45 & 2.48 & 4.00 \\
 & Education & 0.47 & 2.54 & 4.50 \\
\midrule
\multirow{9}{*}{Claude 3.0 Haiku}
 & Age & 0.31 & 3.15 & 5.83 \\
 & Religious & 0.22 & 3.02 & 5.50 \\
 & Socio-Economic & 0.25 & 2.89 & 5.17 \\
 & Ethnicity & 0.30 & 3.02 & 5.50 \\
 & Gender & 0.33 & 2.69 & 4.67 \\
 & Political & 0.20 & 3.36 & 6.83 \\
 & Interest & 0.33 & 2.64 & 4.50 \\
 & Degree-Exclusive & 0.47 & 2.50 & 4.00 \\
 & Education & 0.31 & 2.77 & 5.17 \\
\midrule
\multirow{9}{*}{Llama-4-Scout}
 & Age & 0.29 & 2.79 & 5.50 \\
 & Religious & 0.29 & 2.85 & 5.00 \\
 & Socio-Economic & 0.29 & 2.89 & 5.17 \\
 & Ethnicity & 0.35 & 2.70 & 5.00 \\
 & Gender & 0.33 & 2.72 & 5.00 \\
 & Political & 0.33 & 3.00 & 5.17 \\
 & Interest & 0.31 & 2.68 & 5.00 \\
 & Degree-Exclusive & 0.46 & 2.52 & 4.33 \\
 & Education & 0.24 & 2.83 & 5.00 \\
\bottomrule
\end{tabular}
}
\end{table}

\newpage
\section{Model Convergence Toward Highest-Degree Attachment}
\addcontentsline{toc}{section}{S.I.4. Model Convergence Toward Highest-Degree Attachment}

\begin{figure}[htbp]
    \centering
    \includegraphics[width=0.8\textwidth]{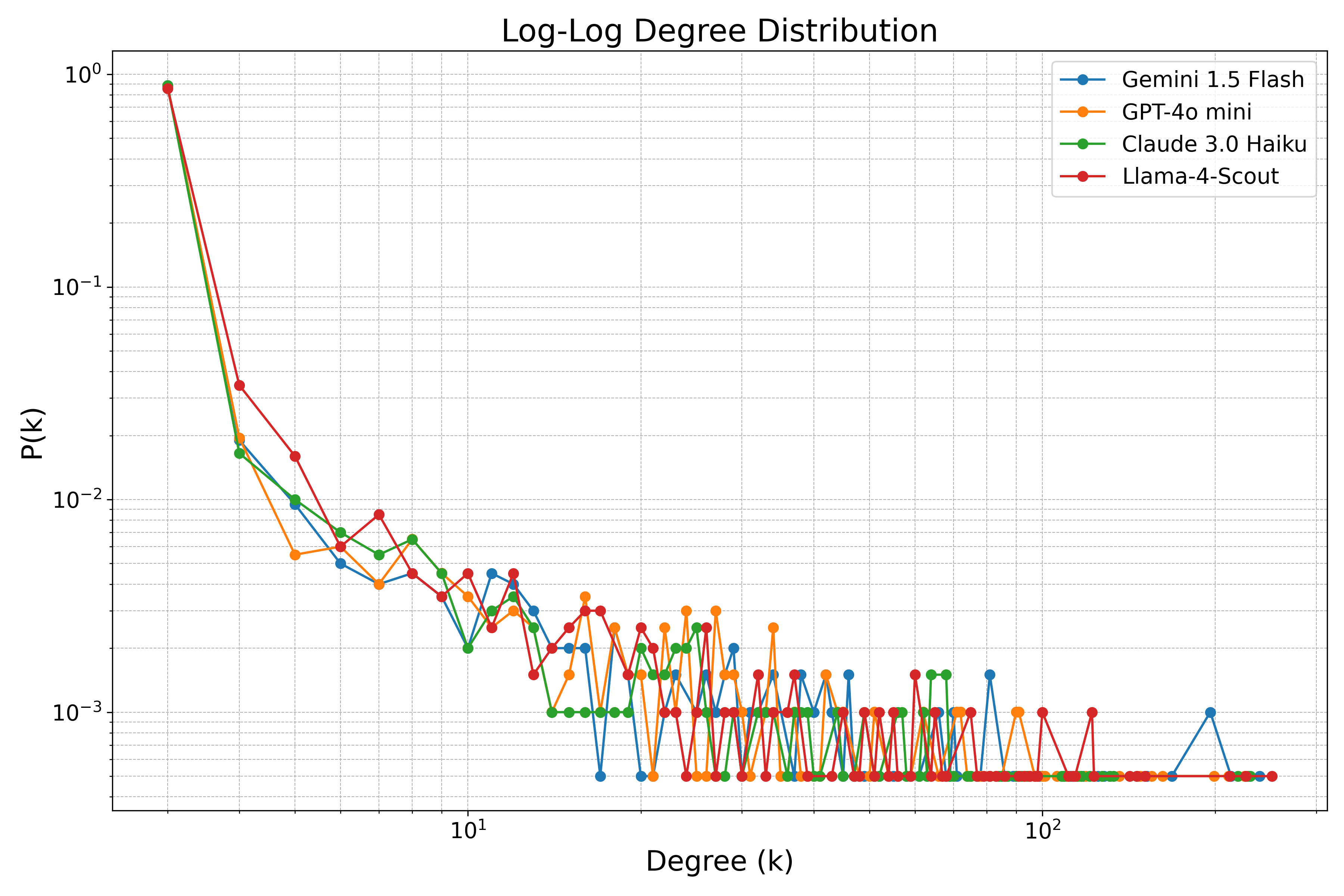}
    \caption{Comparison of degree distributions for networks with 2000 nodes across four LLMs: Gemini 1.5 Flash, GPT-4o Mini, Claude 3.0 Haiku, and Llama-4-Scout. The figure illustrates the close similarity in attachment behavior.}
    \label{fig:model_follow_HDA}
\end{figure}

As observed in Figure~\ref{fig:degree_distributions}, Gemini 1.5 Flash closely follows the highest-degree attachment strategy in the degree-exclusive scenario. To examine whether this behavior generalizes across models, Figure~\ref{fig:model_follow_HDA} compares degree distributions for networks with 2000 nodes generated by four different LLMs: Gemini 1.5 Flash, GPT-4o Mini, Claude 3.0 Haiku, and Llama-4-Scout. Despite differences in architecture, all models exhibit remarkably similar patterns, closely aligning with the highest-degree attachment model. This consistency suggests that when only degree information is available, LLMs tend to default to linking with the most connected nodes reinforcing the dominance of the highest-degree attachment strategy across generative AI systems.

\newpage
\section{S.I.5. Additional Tie Formation Graphs}
\addcontentsline{toc}{section}{S.I.5. Additional Tie Formation Graphs}

\begin{figure}[h!]
    \centering
    \includegraphics[width=0.9\textwidth]{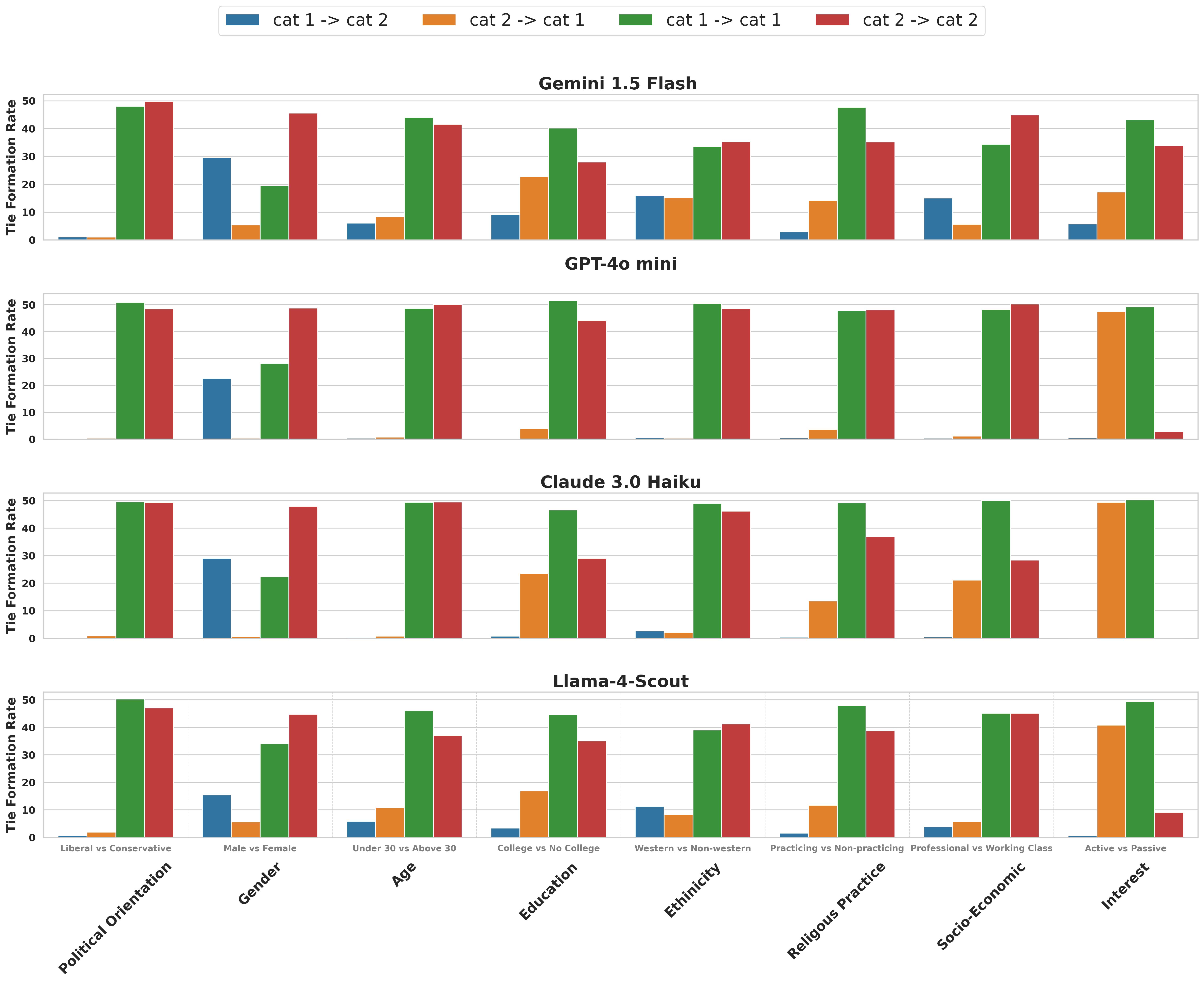}
    \caption{S.I.6: Tie formation rates in LLM-generated networks across four models (Gemini 1.5 Flash, GPT-4o Mini, Claude 3.0 Haiku, and Llama-4-Scout) for binary attributes: political orientation, gender, age, education, ethnicity, religious practice, socio-economic status, and interest. The rates are categorized as cat 1 $\to$ cat 2 (blue), cat 2 $\to$ cat 1 (orange), cat 1 $\to$ cat 1 (green), and cat 2 $\to$ cat 2 (red), highlighting homophily patterns with $m = 3$, $s = 50$.}
    \label{fig:connection_distribution_grid}
\end{figure}

Figure~\ref{fig:connection_distribution_grid} illustrates the directional tie formation patterns across four LLMs for various binary attributes. Homophilous connections, where nodes connect to others with the same attribute, are consistently dominant across all models, with cat 1 $\to$ cat 1 and cat 2 $\to$ cat 2 rates clearly exceeding cross-group links. The symmetry between cat 1 and cat 2 varies by attribute, with attributes like gender and education showing notable asymmetries (e.g., female $\to$ female ties being more frequent than male $\to$ male). This suggests that different LLMs may prioritize similarity differently depending on the social salience of the attribute.

\section{Correlation Between Node Attributes and Degree Centrality}
\label{app:attribute_degree_correlation}

We conducted an additional analysis to investigate whether the micro-level asymmetries observed in tie formation translate into macro-level structural inequalities. To explore this potential effect, we focused our analysis on two attributes that exhibited strong, asymmetric tie formation rates in our primary results: \textit{Gender} and \textit{Religious Practice}. For these two cases, we computed the average final degree for nodes in each category and performed an independent samples t-test to determine if the observed differences were statistically significant.

Our results for these attributes confirm that the group acting as a preferential target for incoming ties consistently accumulates a higher degree. 

\begin{figure}[h!]
    \begin{minipage}{0.48\textwidth}
        \centering
        \includegraphics[width=\linewidth]{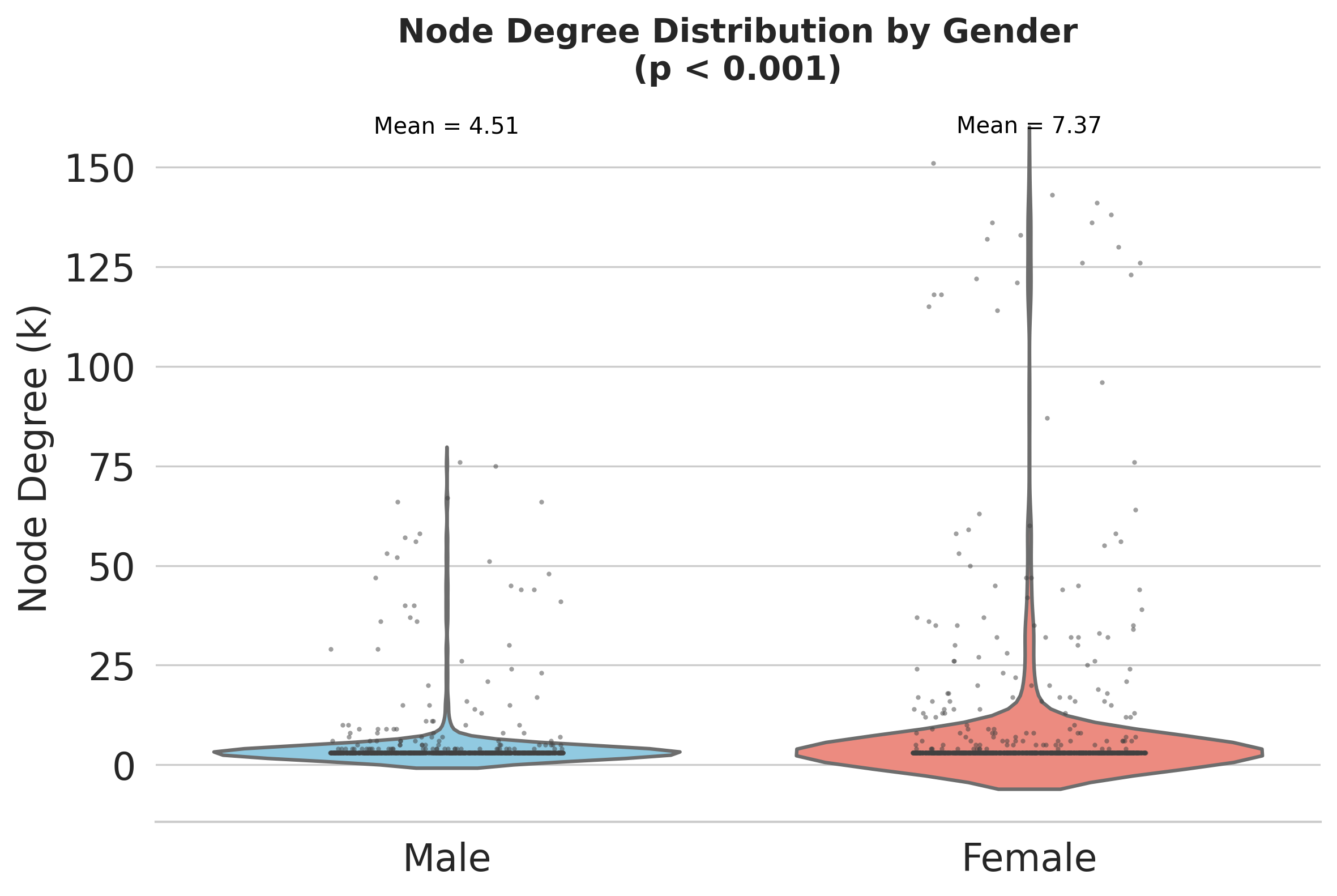}
        \caption{Distribution of final node degrees for the Gender attribute, aggregated across six networks of size $n = 350$. Female nodes achieved a significantly higher average degree than Male nodes.}
        \label{fig:app_degree_dist_gender}
    \end{minipage}\hfill
    \begin{minipage}{0.48\textwidth}
        \centering
        \includegraphics[width=\linewidth]{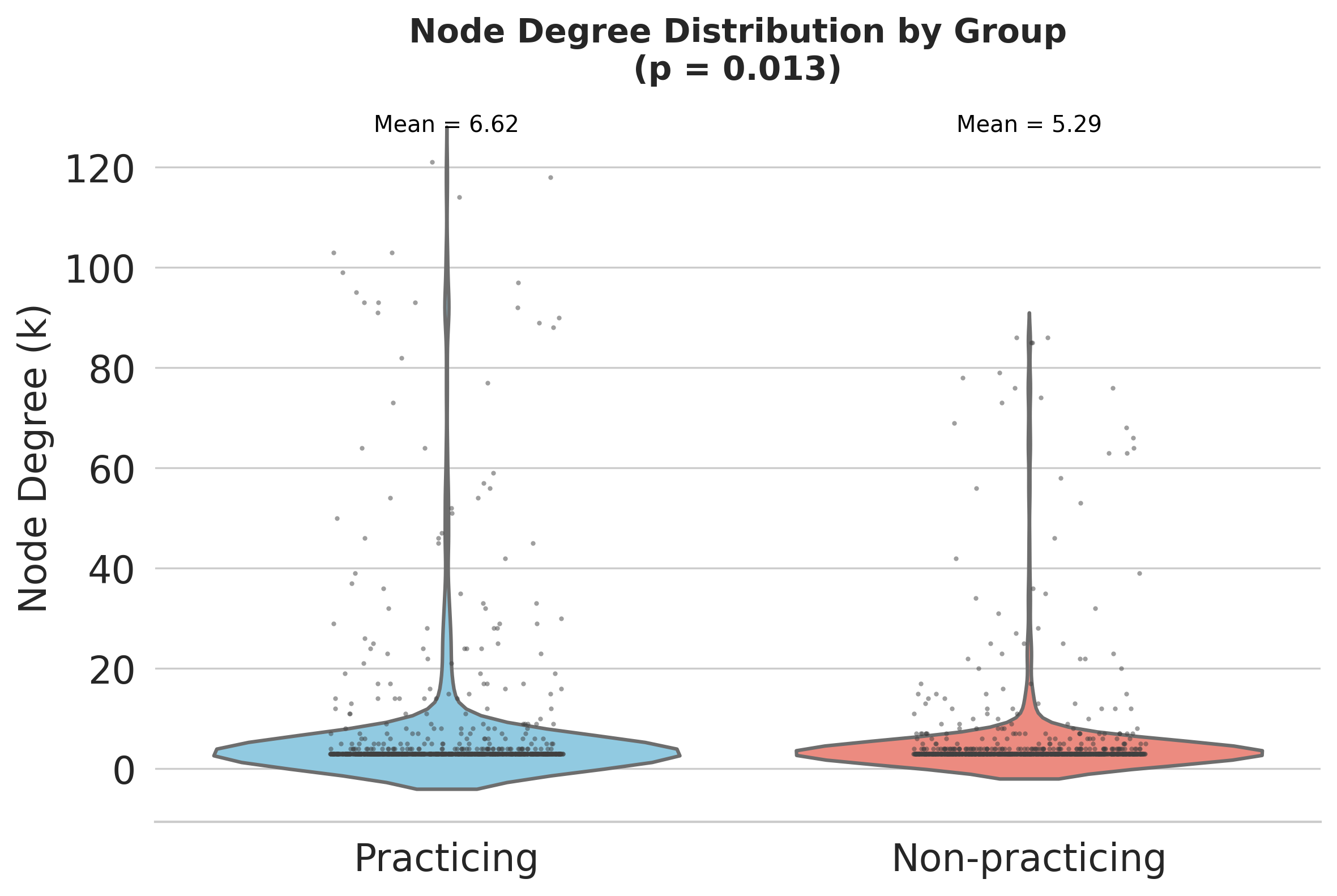}
        \caption{Distribution of final node degrees for the Religious Practice attribute, aggregated across six networks of size $n = 350$. ``Practicing'' nodes accumulated a slightly but statistically significantly higher average degree than ``Non-practicing'' nodes.}
        \label{fig:app_degree_dist_religious}
    \end{minipage}
\end{figure}

For instance, in the ``Gender'' simulations (see Figure \ref{fig:app_degree_dist_gender}), where ``male $\rightarrow$ female'' ties were far more prevalent than the reverse, the average degree for female nodes ($Mean = 7.37$) was significantly higher than for male nodes ($Mean = 4.51$, $p < 0.001$). 

A similar, though less pronounced, effect was observed for ``Religious Practice'' (see Figure \ref{fig:app_degree_dist_religious}). The asymmetry in tie formation, where ``non-practicing $\rightarrow$ practicing'' ties were more frequent, resulted in a statistically significant, albeit smaller, difference in centrality. ``Practicing'' nodes achieved a higher average degree ($Mean = 6.62$) compared to ``Non-practicing'' nodes ($Mean = 5.29$, $p = 0.013$). While this analysis was not exhaustively performed for all attributes, these findings provide preliminary evidence that the directional social biases embedded in LLMs not only affect average node degree but also shape the overall degree distribution, resulting in lasting structural consequences.

\end{document}